\documentclass{99-Styles/CCAP}
\usepackage{afterpage}
\usepackage{longtable}
\usepackage{rotating}
\usepackage{textcomp}
\usepackage{lineno}
\usepackage{tikz}
\usepackage{times}
\usepackage{bm}         
\usepackage{amsmath,amssymb,bm,slashed} 
\usepackage{amssymb}    
\usepackage{graphicx}   
\usepackage{verbatim}
\usepackage{color}      
\definecolor{RedViolet}{cmyk}{0.60, 0.99, 0.99, 0.0}
\definecolor{BlueViolet}{cmyk}{0.90, 0.90, 0.0, 0.0}
\definecolor{DarkBlue}{cmyk}{0.90, 0.65, 0.0, 0.0}
\definecolor{DarkGreen}{cmyk}{0.95, 0.25, 0.95, 0.25}
\definecolor{DarkYellow}{cmyk}{0.0, 0.25, 0.95, 0.25}
\usepackage{subfigure}  
\usepackage{hyperref}   
\usepackage{longtable}
\usepackage{enumerate}
\usepackage{overpic}

\usepackage{fmtcount}   

\usepackage[square,comma,numbers,sort&compress]{natbib}

\usepackage[detect-all]{siunitx}  
\usepackage{enumitem}
\usepackage{blindtext}
\usepackage{wrapfig}
\usepackage{tabularx}
\usepackage{multirow}

\newcommand{\parenbar}[1]{\ensuremath{\overset{\scriptscriptstyle{(-)}}{#1}}}

\newcommand{\lsim}      {\mbox{\raisebox{-0.4ex}{$\;\stackrel{<}{\scriptstyle \sim}\;$}}}

\usepackage{bibentry}
\nobibliography*

\begin{document}

\makeatletter

\newcommand{\bra}[1]{\ensuremath{\langle #1 |}}   
\newcommand{\ket}[1]{\ensuremath{| #1 \rangle}}   
\newcommand{\bigbra}[1]{\ensuremath{\big\langle #1 \big|}}   
\newcommand{\bigket}[1]{\ensuremath{\big| #1 \big\rangle}}   
\newcommand{\amp}[3]{\ensuremath{\left\langle #1 \,\left|\, #2%
                     \,\right|\, #3 \right\rangle}}  
\newcommand{\sprod}[2]{\ensuremath{\left\langle #1 |%
                     #2 \right\rangle}}  
\newcommand{\ev}[1]{\ensuremath{\left\langle #1 %
                     \right\rangle}} 
\newcommand{\ds}[1]{\ensuremath{\! \frac{d^3#1}{(2\pi)^3 %
                     \sqrt{2 E_\vec{#1}}} \,}} 
\newcommand{\dst}[1]{\ensuremath{\! %
                     \frac{d^4#1}{(2\pi)^4} \,}} 
\newcommand{\tr}{\text{tr}}
\newcommand{\sgn}{\text{sgn}}
\newcommand{\diag}{\text{diag}}
\newcommand{\BR}{\text{BR}}

\renewcommand{\vec}[1]{{\mathbf{#1}}}
\renewcommand{\Re}{{\text{Re}}}
\renewcommand{\Im}{{\text{Im}}}
\newcommand{\iso}[2]{{\ensuremath{{}^{#2}}\ensuremath{\rm #1}}}
\newcommand{\eps}{{\ensuremath{\epsilon}}}
\newcommand{\draftnote}[1]{{\bf\color{red} \MakeUppercase{#1}}}
\newcommand{\panm}[1]{{\color{blue} #1}}
\providecommand{\abs}[1]{\lvert#1\rvert}
\providecommand{\norm}[1]{\lVert#1\rVert}

\def\parenbar{\mathpalette\p@renb@r}
\def\p@renb@r#1#2{\vbox{%
  \ifx#1\scriptscriptstyle \dimen@.7em\dimen@ii.2em\else
  \ifx#1\scriptstyle \dimen@.8em\dimen@ii.25em\else
  \dimen@1em\dimen@ii.4em\fi\fi \offinterlineskip
  \ialign{\hfill##\hfill\cr
    \vbox{\hrule width\dimen@ii}\cr
    \noalign{\vskip-.3ex}%
    \hbox to\dimen@{$\mathchar300\hfil\mathchar301$}\cr
    \noalign{\vskip-.3ex}%
    $#1#2$\cr}}}

%
\providecommand{\anmne}{\mbox{$\bar\nu_{\mu} \rightarrow \bar\nu_e$}} 
\providecommand{\nmne}{\mbox{$\nu_{\mu}\rightarrow\nu_e$}} 
\providecommand{\anm}{\mbox{$\bar\nu_\mu$}} 
\providecommand{\nm}{\mbox{$\nu_\mu$}}
\providecommand{\nue}{\mbox{$\nu_e$}} 
\providecommand{\ane}{\mbox{$\bar\nu_e$}} 
\providecommand{\enu}{\mbox{$E_\nu$}}
\providecommand{\piz}{\mbox{$\pi^0 $}}
\providecommand{\pip}{\mbox{$\pi^+$}} 
\providecommand{\pim}{\mbox{$\pi^-$}}

\parindent 10pt
\pagenumbering{roman}
\setcounter{page}{1}
\pagestyle{plain}

\thispagestyle{empty}

\leftline{} \vspace{-0.4cm} \rightline{March 31, 2025} \vspace{0.5cm}

\noindent{\bf\LARGE 
  Neutrinos from Stored Muons (nuSTORM) \\
}
\noindent\textit{Submitted to the 2026 Update of the European Strategy for Particle Physics.}

\vspace{0.25cm}
\noindent \textbf{Contact: K.~Long}

\renewcommand{\thefootnote}{\fnsymbol{footnote}}
\noindent
Prepared on on behalf of the nuSTORM collaboration\footnote[2]{
The nuSTORM collaboration is presented in~\cite{nuSTORMsnow}.
} 
by:\\
\noindent
L.~Alvarez~Ruso${}^{1}$,
W.~Chang${}^{2,3}$,
J.~Franklin${}^{4}$,
P.R.~Hobson${}^{5}$,
P.B.~Jurj${}^{6}$,    
R.~Kamath${}^{6,3}$,
P.~Kyberd${}^{7}$,
X.~Lu${}^{2}$
K.~Long${}^{6,3}$,
D.~Pasari${}^{4}$,
S.~Ricciardi${}^{3}$,
J.~Turner${}^{4}$,
A.~Vogiatzi${}^{6}$
\begin{flushleft}
  \begin{minipage}{17cm}
    {\em\footnotesize
      ${}^{1}$ Instituto de Física Corpuscular, Parque Científico,
        Catedrático José Beltrán, 2, E-46980 Paterna, Spain \\
      ${}^{2}$ Department of Physics, University of Warwick, Coventry,
        CV4 7AL, UK \\
      ${}^{3}$ STFC, Rutherford Appleton Laboratory, Harwell Campus,
       Didcot, OX11 0QX  \\
      ${}^{4}$ Department of Physics, The Palatine Centre, Durham University,
        Stockton Road, Durham, DH1 3LE, UK \\
      ${}^{5}$ Department of Physics and Astronomy, Queen Mary University
        of London, London, E1 4NS, UK \\
      ${}^{6}$ Physics Department, Blackett Laboratory, Imperial
        College London, Exhibition Road, London, SW7 2AZ, UK \\
      ${}^{7}$ College of Engineering, Design and Physical Sciences,
        Brunel University, Uxbridge, Middlesex, UB8 3PH, UK \\
    }
  \end{minipage}
\end{flushleft}

\vspace{-0.20cm}
  The Neutrinos from Stored Muons, nuSTORM, facility has been designed
  to deliver a definitive neutrino-nucleus scattering programme using
  beams of $\parenbar{\nu}\!\!_e$ and $\parenbar{\nu}\!\!_\mu$ from the decay
  of muons confined within a storage ring.
  The facility is unique, it will be capable of storing $\mu^\pm$
  beams with a central momentum of between 1\,GeV/c and 6\,GeV/c and a
  momentum spread of 16\%. 
  This specification will allow neutrino-scattering measurements to be made
  over the kinematic range of interest to the DUNE and Hyper-K collaborations.
  At nuSTORM, the flavour composition of the beam and the
  neutrino-energy spectrum are both precisely known.
  The storage-ring instrumentation will allow the neutrino flux to be
  determined to a precision of 1\% or better.
  By exploiting sophisticated neutrino-detector techniques such as
  those being developed for the near detectors of DUNE and Hyper-K,
  the nuSTORM facility will:
  \begin{itemize}[noitemsep,topsep=0pt]
    \item Serve the future long- and short-baseline neutrino-oscillation
      programmes by providing definitive measurements of
      $\parenbar{\nu}\!\!_e A$ and $\parenbar{\nu}\!\!_\mu A$
      scattering cross-sections with percent-level precision;
    \item Provide a probe that is 100\% polarised and sensitive to
      isospin to allow incisive studies of nuclear dynamics and
      collective effects in nuclei;
    \item Deliver the capability to extend the search for light
      sterile neutrinos beyond the sensitivities that will be provided
      by the FNAL Short Baseline Neutrino (SBN) programme; and
    \item Through the delivery of a unique neutrino-physics programme,
      create an essential test facility for the development of muon
      accelerators to serve as the basis of a multi-TeV
      lepton-antilepton collider. 
  \end{itemize}
  To maximise its impact, nuSTORM should be implemented such that
  data-taking coincides with the accumulation of substantial data
  samples by the the DUNE and Hyper-K collaborations.
  This will allow measurements at nuSTORM to be used to resolve the
  correlation between flux and cross-section uncertainties that
  naturally arise in the analysis of near-detector data and so allow
  oscillation probabilities to be determined with percent-level
  precision. 

  With its existing proton-beam infrastructure, CERN is uniquely
  well-placed to implement nuSTORM.
  The feasibility of implementing nuSTORM at CERN has been studied by
  a CERN Physics Beyond Colliders study group~\cite{Ahdida:2020whw}.
  The muon storage ring has been optimised for the neutrino-scattering
  programme to store muon beams with momenta in the range 1\,GeV/c to 
  6\,GeV/c.
  The implementation of nuSTORM exploits the existing fast-extraction
  from the SPS that delivers beam to the LHC and to HiRadMat.
  A summary of the proposed implementation of nuSTORM at CERN is
  presented below.
  An indicative cost estimate and a preliminary discussion of a 
  possible time-line for the implementation of nuSTORM are presented 
  in~\cite{Ahdida:2020whw}.

\cleardoublepage
\pagenumbering{arabic}                   
\setcounter{page}{1}

\graphicspath{ {01-Scientific-context/Figures} }

\section{Scientific context}
\label{Sect:SciCon}

The Neutrinos from Stored Muons, nuSTORM, facility has been designed
to provide intense neutrino beams with well-defined flavour
composition and energy spectra.
By using neutrinos from the decay of muons confined within a storage
ring, a beam composed of equal fluxes of electron- and muon-neutrinos
can be created for which the energy spectrum can be calculated
precisely.
The case for the nuSTORM facility rests on three key themes:
\begin{enumerate}[noitemsep,topsep=0pt]
  \item \textbf{Precision Neutrino Scattering Studies}: nuSTORM's
    uniquely well-defined neutrino beam enables precise
    neutrino-nucleus scattering measurements over the energy range
    relevant to long- and short-baseline experiments.
    Future long-baseline experiments such as Hyper-K and DUNE will
    achieve percent-level statistical uncertainties only after a
    few years of running.
    The full exploitation of the statistical weight of the data
    samples will therefore require comparable systematic precision.
    nuSTORM can contribute to maximising the scientific return by
    providing direct cross-section measurements, breaking
    flux-correlation uncertainties, and reducing overall systematics 
    to optimise discovery potential.
  \item \textbf{Beyond Standard Model (BSM) Physics}: The combination
    of nuSTORM’s instrumented neutrino beam with state-of-the-art,
    magnetised near and far detectors allows unprecedented sensitivity
    for BSM searches.
    The resulting signal-to-background ratio is significantly higher
    than in other accelerator-based experiments. 

    For short-baseline sterile-neutrino searches, nuSTORM offers an
    initial flavour composition that is precisely defined which, in
    combination with a precisely known neutrino-energy spectrum,
    ensures minimal background uncertainties.   
  \item \textbf{Muon Accelerator R\&D}: nuSTORM’s stored muon beam
    serves as a test-bed for next-generation high flux/high brightness
    muon accelerators, critical for a future Neutrino Factory and Muon
    Collider.
    By demonstrating key technologies while enabling a unique
    portfolio of neutrino physics, nuSTORM lays the foundation for
    high-energy lepton colliders and intense neutrino sources.  
\end{enumerate}

Together, these three pillars make nuSTORM a uniquely robust facility,
both scientifically productive and instrumental in advancing particle
physics technologies.

\subsection{Neutrino scattering}
\label{SubSect:SciCon:nuSctt}

\graphicspath{ {01-Scientific-context/Figures/} }

nuSTORM provides a unique facility for generating precisely controlled
electron neutrino ($\nu_e$) and electron antineutrino ($\bar{\nu}_e$)
beams from stored muons.
This enables precise measurements of neutrino interactions at
GeV-scale energies, addressing key uncertainties in neutrino
oscillation experiments such as Hyper-Kamiokande (HyperK) and the Deep
Underground Neutrino Experiment (DUNE). 

CP violation in the lepton sector requires three-flavour neutrino
oscillation, which involves the appearance of $\nu_e$ and
$\bar{\nu}_e$ from $\nu_\mu$ and $\bar{\nu}_\mu$ beams.
However, conventional long-baseline neutrino experiments rely on
neutrino beams generated from hadron decays, which primarily produce
$\nu_\mu$ and $\bar{\nu}_\mu$.
The resulting electron (anti)neutrino component is an intrinsic
background; 
 using these $\nu_e$ for precise cross-section measurements is challenging due to their energies, which are not relevant for oscillations, and the significant contamination from $\nu_\mu$ and $\bar{\nu}_\mu$.

A major systematic uncertainty in CP violation measurements is the
$\nu_e$ and $\bar{\nu}_e$ cross-section uncertainty.
Unlike $\nu_\mu$ and $\bar{\nu}_\mu$ interactions, in situ
measurements of $\nu_e$ and $\bar{\nu}_e$ cross sections are not
feasible as argued above.
As a result, CP violation searches depend on theoretical models and
indirect measurements, leading to uncertainties that limit
sensitivity. As an example, for HyperK~\cite{Hyper-KamiokandeProto-:2015xww},
reducing the $\nu_e/\bar{\nu}_e$ cross-section ratio uncertainty from
4.9\% to 2.7\% can improve CP violation sensitivity by 1$\sigma$ with
six years of data collection. 
Alternatively, it could reduce the required data-taking period from
ten years to five to reach a 5$\sigma$ discovery level.
This demonstrates the critical need for dedicated $\nu_e$
cross-section measurements. 

To enable precise CP violation measurements, an ideal $\nu_e$
cross-section experiment must satisfy: 
\begin{itemize}[noitemsep,topsep=0pt]
  \item Well-understood fluxes with controlled uncertainties;
  \item Correct beam energy settings to match oscillation experiments;
  \item High-statistics data collection; 
  \item Low $\nu_\mu$ background to minimise contamination; and
  \item Optimised detector design for $\nu_e$ interactions.
\end{itemize}
nuSTORM provides several unique advantages that address these
requirements:
\begin{itemize}[noitemsep,topsep=0pt]
  \item Optimisable $\nu_e$ and $\bar{\nu}_e$ fluxes with a perfectly
    known flux shape and normalisation;
  \item Precise control over the neutrino beam energy, matching the
    oscillation-relevant regime of HyperK ($\sim$0.6~GeV) and DUNE
    ($\sim$2.4~GeV);
  \item Neutrino beams from stored muons are free from
    hadron-decay-induced backgrounds, e.g. there is no $\bar{\nu}_e$
    background accompanying the $\nu_e$ flux from a stored $\mu^+$
    beam;
  \item Accelerator tuning for fine control over the flux shape and
    spread; 
  \item Polarised neutral lepton beams, allowing for isospin-sensitive
    measurements; and
  \item Unique capability for neutrino beam energy-scan (nuBES)
    measurements to map cross-section energy dependencies and reduce
    systematic errors. 
\end{itemize}

Besides the direct impact on the determination of neutrino properties, the precise measurements in a nuSTORM facility would provide valuable information about the axial structure of nucleons and nuclei. The available information about neutrino-nucleon interactions is scarce and comes mostly from old bubble chamber experiments. These cross sections could be measured directly using hydrogen or deuterium targets or indirectly with the help of hydrogen-enriched target and subtraction techniques. Even for a basic quasielastic process, the dependence of the axial form factor $F_A$ on the four-momentum transferred to the nucleon squared ($Q^2$) is not precisely measured. 
Neutrinos also scatter inelastically on nucleons, predominantly leading to single pion ($\pi N$) but also many other baryon-meson final states. The cross section arises from the interplay of resonant and non-resonant amplitudes, 
with several overlapping resonances and coupled channels. A large fraction of events at DUNE will belong to this category. Progress in the precise description of the the transition towards the deep-inelastic region is hindered by the lack of experimental data.  For heavy targets,  nuSTORM 
can play an important role in understanding discrepancies with theory have been found by NOvA and MINERvA. Finally, the characterization of nuclear corrections to parton distribution functions will also benefit form precise measurements to unravel the differences in nuclear effects observed in weak and electromagnetic processes. 

\noindent\textbf{nuSTORM Neutrino Beam Energy Scan (nuBES) and Nuclear Dynamics} \\
\noindent nuSTORM's ability to scan neutrino energies enables detailed studies
of nuclear dynamics.
One key example is the use of Transverse Kinematic Imbalance
(TKI)~\cite{Lu:2015tcr} techniques to explore nuclear effects in neutrino
interactions.
By systematically varying the neutrino energy, nuSTORM provides
critical data to validate and refine neutrino interaction models
thereby improving the accuracy of long-baseline oscillation
experiments.  
\begin{figure}[!htb]
  \begin{center}
    \includegraphics[width=0.5\linewidth]{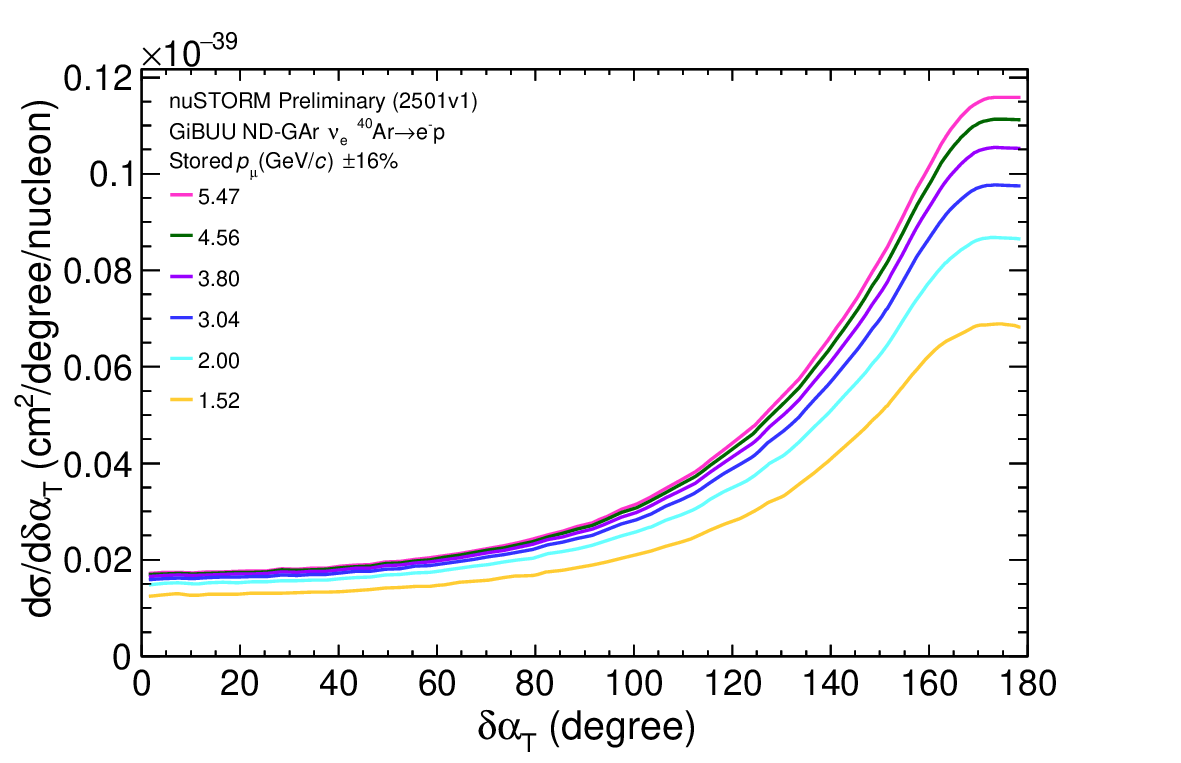}
  \end{center}
  \caption{
    Model predictions of differential cross section in the transverse
    boosting angle, $\delta\alpha_\textrm{T}$, for different  stored
    muon momenta in nuSTORM.
    GiBUU~\cite{Buss:2011mx, Mosel:2023zek} simulation for
    charged-current pionless $\nu_e$ scattering on argon is performed
    in DUNE's ND-GAr detector~\cite{DUNE:2021tad} acceptance.
  }
  \label{fig:nustormtki}
\end{figure}

\noindent\textbf{nuSTORM Neutrino Beam Energy Scan (nuBES) and Synthetic Beams} \\
\noindent
The nuSTORM facility can generate synthetic neutrino beams by
combining samples generated using a number of stored-muon-beam
energies.
The process is similar in concept to the PRISM technique that will be
used at DUNE or HyperK \cite{Bhadra2014LetterBeamlineb}, albeit while keeping the detector fixed
on-axis.
Instead of repositioning the detector, nuSTORM exploits the
energy-dependent variation in neutrino flux—--narrower peaks at lower
energies and broader ones at higher energies.
This enables the use of the full neutrino-detector suite throughout.

The neutrino flux can be expressed as:
$    \Phi_{LC}(E_{\nu}) = \sum_{i}^{N_{\mu}} c_i\phi_i(E_{\nu}) \, $,
where $\phi_i(E_{\nu})$ represents the neutrino energy spectrum at
different muon momenta, and $c_i$ are weighting coefficients.
The coefficients are optimised by minimising the figure of merit (FOM):
\begin{equation} \label{eq:FOM}
    {\rm FOM} = \sum_{E_\nu} \frac{(f(E_\nu) -
                \Phi_{LC}(E_{\nu}))^2}{A + Bf(E_\nu)^2} \,,
\end{equation}
where $f(E_\nu)$ is the target function, which may be a normal
distribution for neutrino-nucleus interactions or an arbitrary
spectrum.
The parameters $A$ and $B$ adjust weighting in the fit. 

\begin{figure}[!h]
  \begin{center}
    \begin{minipage}{0.45\textwidth}
      \begin{center}
        \includegraphics[width=\textwidth]{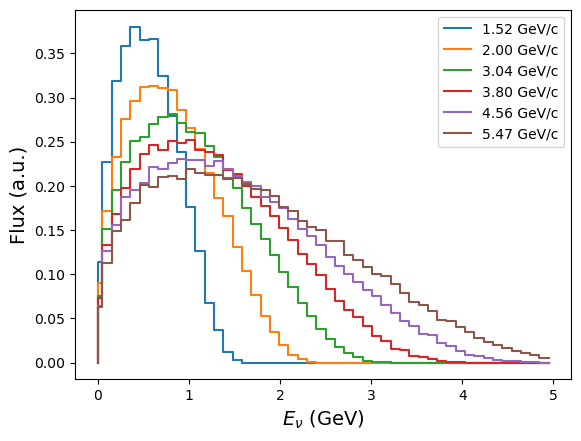}
      \end{center}
    \end{minipage}%
    \begin{minipage}{0.45\textwidth}
      \begin{center}
        \includegraphics[width=\textwidth]{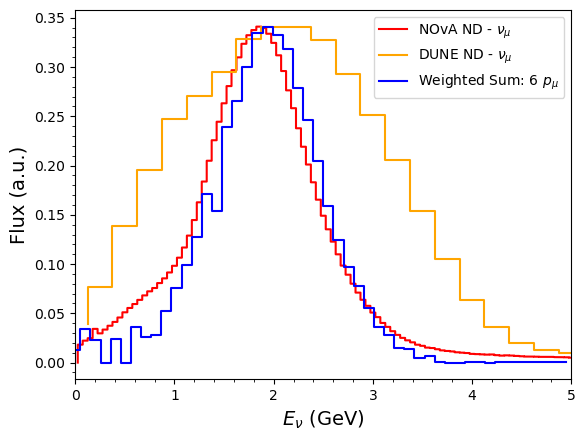}
      \end{center}
    \end{minipage}%
  \end{center}
  \caption{(Left) The flux from six different nuSTORM muon neutrino settings. (Right) a 2 $\pm$ 0.5 GeV reconstructed synthetic neutrino beam, compared with NOvA \cite{Shanahan2021PhysicsReview} and DUNE \cite{DUNE:2021tad}.}
  \label{fig:nuSYNTH}
\end{figure}

As shown in figure~\ref{fig:nuSYNTH}, nuSTORM can generate
quasi-mono-energetic beams with a linear combination of spectra,
achieving a synthetic beam with a 65\% narrower FWHM than the natural
muon decay spectrum~\cite{Kamath2024SimulationExperiment}.
Unlike off-axis techniques like DUNE-PRISM, nuSTORM can also produce
synthetic $\nu_e$ beams by combining different $\nu_e$ spectra.

\graphicspath{ {01-Scientific-context/Figures/} }

\subsection{Searching for Physics Beyond the Standard Model}
\label{SubSect:SciCon:BSM}

nuSTORM has the capability to test a number of new physics
scenarios~\cite{Chakraborty_2021}, including the existence of large
extra dimensions
(LED) \cite{Arkani-Hamed:1998jmv,Arkani-Hamed:1998sfv} which offer a
solution to the hierarchy problem by embedding our $3+1$‑dimensional
universe (the “brane”) within a higher‑dimensional bulk. 
Sterile neutrinos, being neutral under Standard Model charges, can
traverse the bulk—compactified on circles of radius
$R_{\mathrm{LED}}$—while all other particles remain confined to the
brane.
From a four‑dimensional perspective, each bulk neutrino appears as an
infinite Kaluza–Klein tower.
Yukawa couplings between these modes, brane‑localised left‑handed
neutrinos and the Higgs boson generate Dirac masses suppressed by the
extra‑dimensional volume, naturally explaining the smallness of
neutrino masses.
Mixing between active neutrinos and the Kaluza–Klein states modifies
oscillation probabilities, allowing precision neutrino experiments to
constrain the size of the extra dimension.

The phenomenology is similar to that of light sterile neutrinos, which
induce observable short-baseline neutrino oscillations and small
perturbations to the neutrino oscillations in solar, atmospheric and
long-baseline experiments that are well-described by standard
three-neutrino mixing. 
In the LED framework, one begins with an effective Lagrangian that
couples the three brane‑confined left‑handed neutrinos to an infinite
tower of bulk Kaluza–Klein modes.
Diagonalising the resulting mass matrix yields the mass eigenvalues
and mixing coefficients that determine how each flavour state projects
onto the KK modes.
The survival probability is then calculated by summing over all mass
eigenstates—each contributing a phase factor that depends on its
squared mass, travel distance and energy.

\begin{figure}[h]
  \begin{center}
    \includegraphics[width=0.40\textwidth]{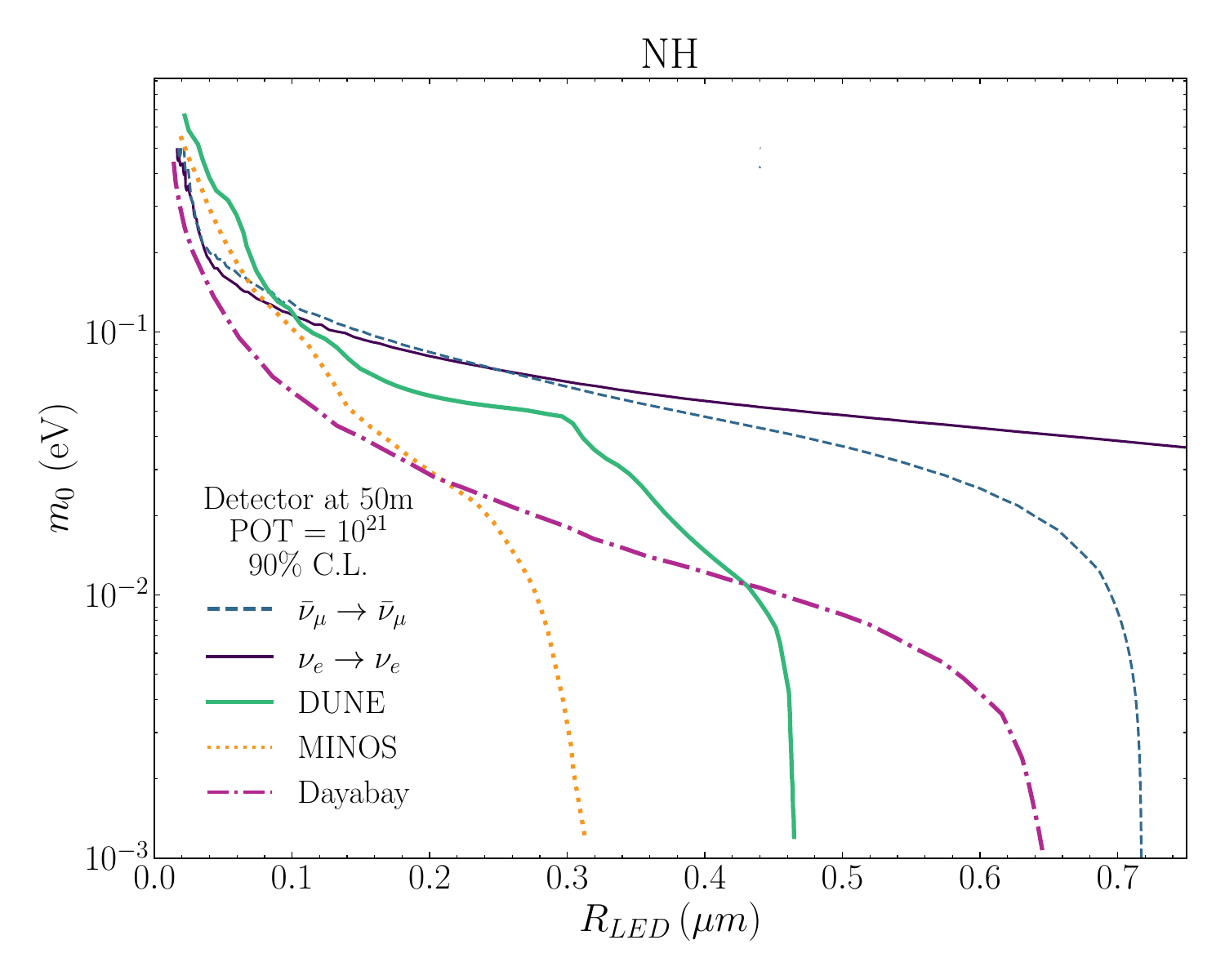}
    \includegraphics[width=0.40\textwidth]{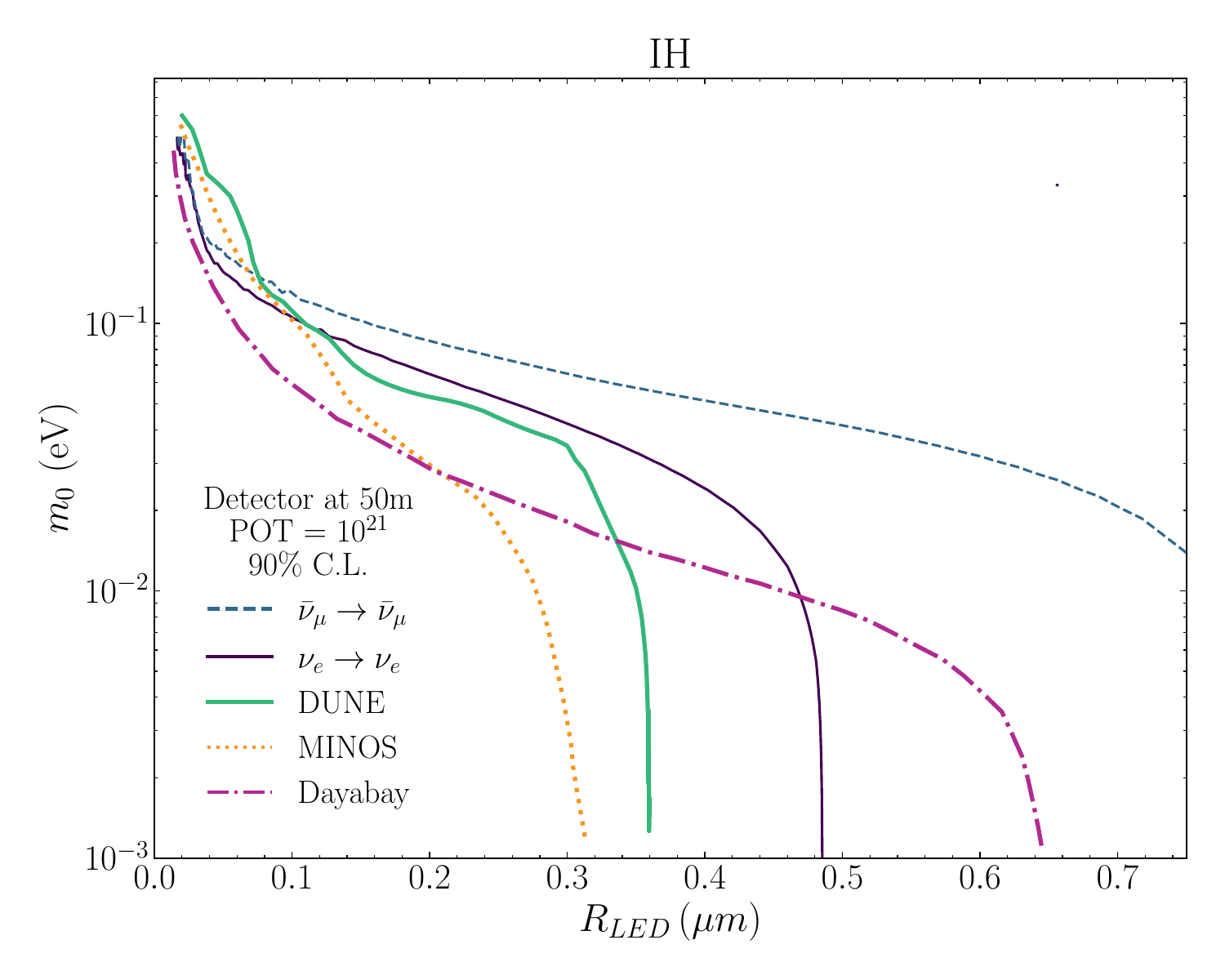}
  \end{center}
  \caption{
    Sensitivity to disappearance under the assumption of 1\%
    systematic uncertainty for the normal hierarchy (left) and
    inverted hierarchy (right).
    Comparison against contours from DUNE, MINOS, and Daya Bay is also
    shown.
  }
  \label{fig2:chi_square_led_oscillation}
\end{figure}

By truncating the KK tower numerically and performing a $\chi^2$
comparison of the predicted oscillation probabilities with
experimental data assuming $1\%$ systematic uncertainties, we obtain
constraints on the lightest Dirac neutrino mass, $m_0$, and the radius
of the extra dimension shown in
figure~\ref{fig2:chi_square_led_oscillation} which shows the $90\%$ 
(solid) confidence limit for two degrees of freedom.
In the Normal Hierarchy (NH) scenario, nuSTORM is projected to exclude
\(R_{\mathrm{LED}}\lesssim0.1\,\mu\mathrm{m}\) for a lightest Dirac
mass \(m_0\gtrsim0.1\,\mathrm{eV}\), providing a modest improvement
over the current bounds from DUNE and MINOS; for
\(m_0<0.1\,\mathrm{eV}\), however, existing constraints from MINOS,
Daya Bay and DUNE remain more restrictive. Under the Inverted
Hierarchy (IH) assumption, nuSTORM likewise yields slightly stronger
exclusion limits than DUNE and MINOS in the high‑mass regime
(\(m_0\gtrsim0.1\,\mathrm{eV}\)), and—unlike the NH case—its
sensitivity surpasses that of Daya Bay for \(m_0<0.1\,\mathrm{eV}\). 

\begin{figure}[h]
  \begin{center}
    \includegraphics[width=0.40\linewidth]{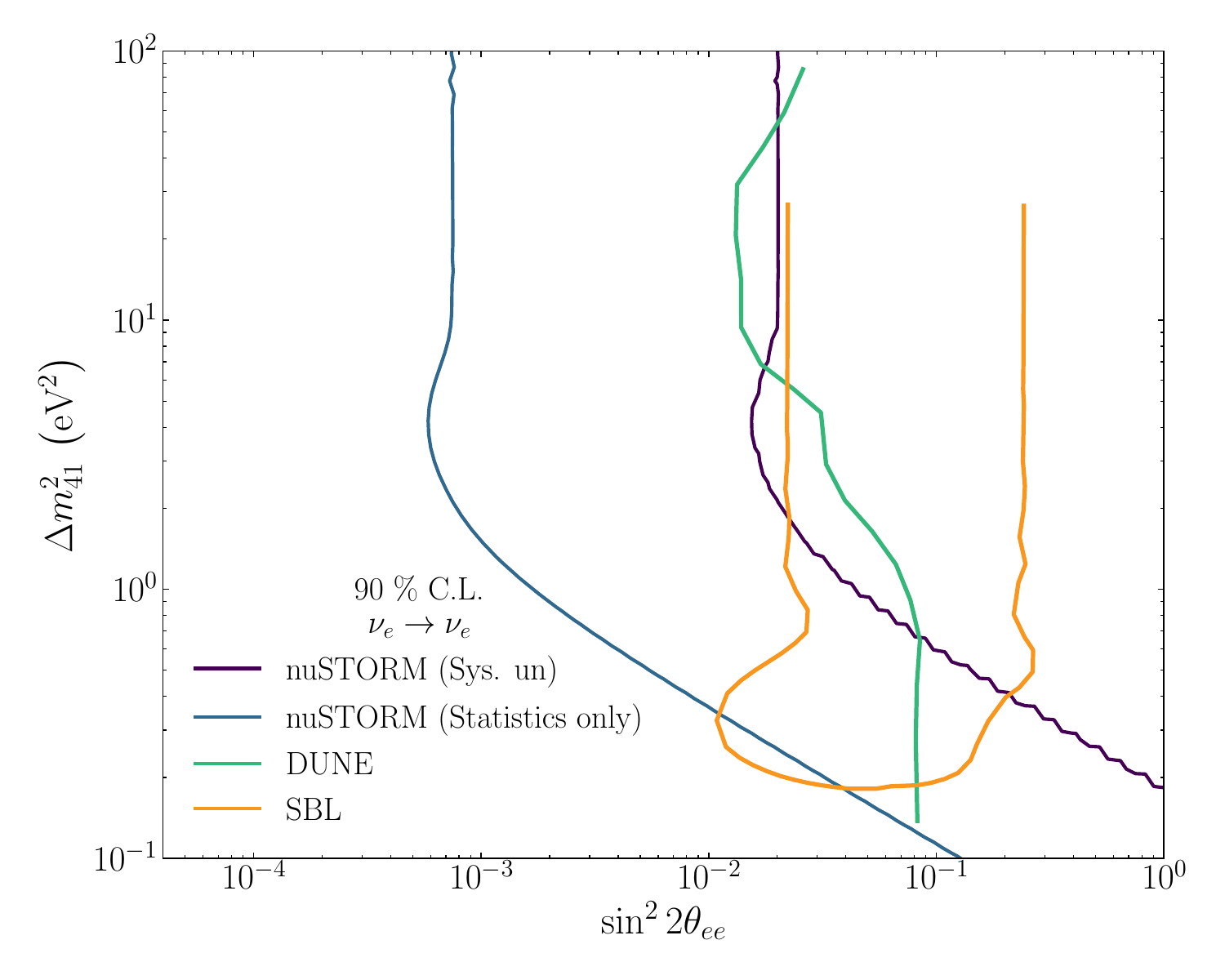}
    \includegraphics[width=0.40\linewidth]{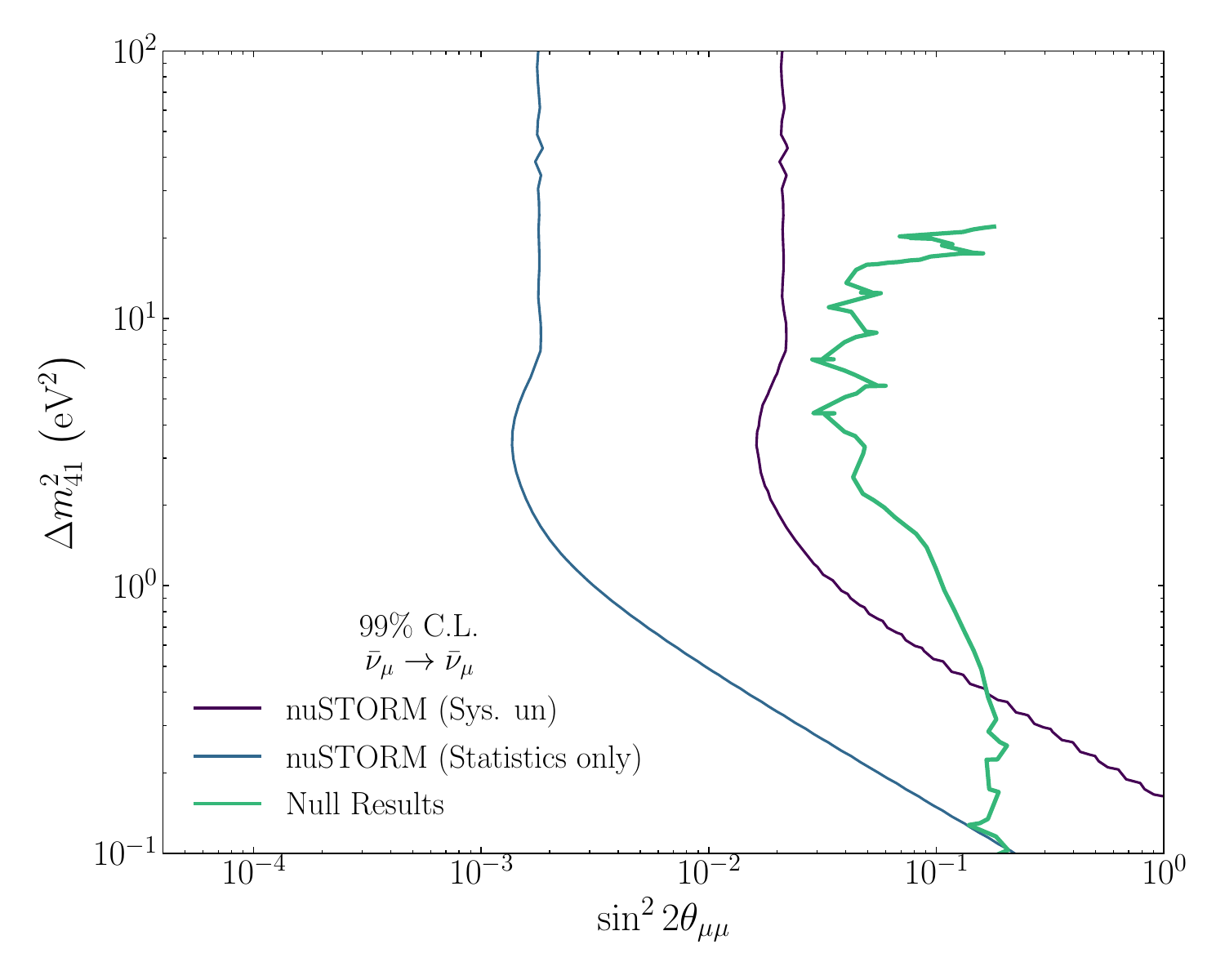}
  \end{center}
  \caption{
    Sensitivity to disappearance under the assumption of 1\%
    systematic uncertainty.
    Comparison against contours from other short‑baseline experiments
    is shown.
    The SBL reactors used for this analysis are Bugey3
    \cite{Declais:1994su}, Bugey4 \cite{Declais:1994ma}, Gosgen
    \cite{CALTECH-SIN-TUM:1986xvg}, ILL \cite{Kwon:1981ua},
    Krasnoyarsk \cite{Vidyakin:1987ue}, SRP \cite{Greenwood:1996pb},
    Rovno91 \cite{Kuvshinnikov:1990ry}, and Rovno88
    \cite{Afonin:1988gx}, compiled in \cite{Kopp:2013vaa}.
    The combined null result from MINOS \cite{MINOS:2011ysd}, CDHS
    \cite{Dydak:1983zq}, Super‑Kamiokande \cite{Bilenky:1999ny}, and
    MiniBooNE \cite{MiniBooNE:2012meu} (also compiled in
    \cite{Kopp:2013vaa}) is used for the \(\nu_\mu\) channel.
  }
  \label{fig:chi_square_sterile_neutrino}
\end{figure}

The nuSTORM facility, which features high‑intensity,
well-characterised neutrino beams in which, for a particular neutrino
flavour, there is no contamination from the charge-conjugate state,
and for which systematic uncertainties will be at the sub-percent
level, is uniquely positioned to probe the 3+1 sterile‑neutrino
parameter space. 
In the short‑baseline approximation, the disappearance probability for
flavour \(\alpha\) in a 3+1 framework is
\begin{equation}
P_{\alpha\alpha}^{\mathrm{SBL}} = 1 -
\sin^2(2\theta_{\alpha\alpha})\,\sin^2\!\biggl(\frac{\Delta
  m^2_{41}L}{4E}\biggr) \, ;
\end{equation}
where \(\Delta m^2_{41}\) and \(\theta_{\alpha\alpha}\) denote the
mass splitting and effective mixing angle between the active and
sterile states \cite{Penedo:2022etl,Kopp:2013vaa}. 
To quantify nuSTORM’s sensitivity to sterile‑neutrino disappearance,
we perform a \(\chi^2\) analysis comparing the predicted disappearance
probability against the null‑oscillation hypothesis.
Neutrino production in the muon decay straight is modelled by
uniformly sampling baselines in the range \(50\text{ m} \le L \le
250\text{ m}\).
The resulting exclusion contours are presented in
Fig.~\ref{fig:chi_square_sterile_neutrino}.

In the \(\bar\nu_\mu\) disappearance channel, nuSTORM achieves
substantially stronger limits than the current combined null results
for \(\Delta m^2_{41} \gtrsim 0.3\,\mathrm{eV}^2\).
Likewise, in the \(\nu_e\) channel, nuSTORM outperforms existing
short‑baseline reactor experiments and the projected sensitivity of
DUNE over the range \(2\,\mathrm{eV}^2 \lesssim \Delta m^2_{41}
\lesssim 6\,\mathrm{eV}^2\).
Furthermore, nuSTORM's magnetised liquid‑argon time projection chamber
(LArTPC) provides robust charge identification for \(\nu_\mu\) and
\(\bar\nu_\mu\) events, enabling independent measurements of
disappearance probabilities in both neutrino and antineutrino modes.
This capability offers a unique avenue to probe potential CP‑violating
effects in the sterile‑neutrino sector. 

nuSTORM provides an ideal platform to search for charged‑lepton
flavour violation (cLFV) in rare meson decays.
Observation of neutrino oscillations demonstrates that lepton flavour
is not conserved, violating an accidental symmetry of the Standard
Model.
Charged‑lepton flavour–violating processes remain exceedingly
suppressed within the Standard Model, owing to the tiny neutrino
masses, and any measurable signal would constitute unambiguous
evidence for physics beyond the Standard Model
\cite{Kuno:1999jp,Bernstein:2013hba}.
Numerous theoretical frameworks, including super-symmetric models,
leptoquark scenarios, and neutrino‑mass generation mechanisms, predict
significantly enhanced cLFV rates
\cite{Lindner:2016bgg,Calibbi:2017uvl}, motivating increasingly
stringent experimental constraints.
Historically, the most sensitive cLFV searches have targeted rare muon
decays (\(\mu\to e\gamma\) and \(\mu\to e\) conversion in nuclei),
yielding upper limits on branching ratios at the
\(10^{-13}\)–\(10^{-15}\) level \cite{MEG:2016leq,Bernstein:2019fyh}.
Complementary limits arise from meson leptonic decays, the Big
European Bubble Chamber (BEBC) experiment set the current best
constraint on the lepton‑flavour‑violating pion decay
$\mathrm{BR}(\pi^+\to\mu^+\nu_e)<8\times10^{-3}$ compiled in
references~\cite{Cooper-Sarkar:1981bam,Lyons:1981xs,ParticleDataGroup:2022pth},
with further improvements anticipated in future studies
\cite{Alves:2024djc}.
nuSTORM’s high flux of well‑characterised pions enables a direct
search for \(\pi^+\to\mu^+\nu_e\).
The expected signal yield is:
\begin{equation}
  N_{\nu_e,\mathrm{sig}} = N_{\nu_\mu}^{\pi^+}\,\mathrm{BR}(\pi^+\to\mu^+\nu_e) \,;
\end{equation}
where \(N_{\nu_\mu}^{\pi^+}\) denotes the number of muon neutrinos
produced via the Standard Model decay channel. Sensitivity is
evaluated via a \(\chi^2\) statistic:
\begin{equation}
  \chi^2 = \frac{N_{\nu_e,\mathrm{sig}}^2}{N_{\nu_e} +
                             \sigma_{\nu_e}^2\,N_{\nu_e}^2}\,;
\end{equation}
with \(N_{\nu_e}\) the intrinsic electron‑neutrino background and
\(\sigma_{\nu_e}\) its systematic uncertainty \cite{Alves:2024djc}.
Table~\ref{tab:LFV} summarises the projected nuSTORM limits relative
to existing constraints.
nuSTORM improves the upper limit on
\(\mathrm{BR}(\pi^+\to\mu^+\nu_e)\) by approximately a factor of 11
compared to BEBC and by a factor of 2 compared to SBND.
Although nuSTORM's intrinsic background ratio
\((\nu_e:\nu_\mu)_{\mathrm{nuSTORM}}\approx0.036\) exceeds those of
BEBC
\((\nu_e:\nu_\mu)_{\mathrm{BEBC}}\approx0.016\)\cite{Cooper-Sarkar:1981bam}
and SBND
\((\nu_e:\nu_\mu)_{\mathrm{SBND}}\approx0.008\)\cite{Alves:2024djc},
its vastly larger sample of muon‑neutrino charged‑current events —
\(6060\pm440\) at BEBC, \(\mathcal{O}(10^7)\) at SBND, and
\(\mathcal{O}(10^8)\) at nuSTORM — yields substantially greater
statistical power.
This dramatic increase in signal statistics more than compensates for
the higher background, enabling nuSTORM to set the most stringent
constraint to date on this lepton‑flavour‑violating decay. 
\begin{table}[h]
  \caption{
    Comparing the bounds on LFV against bounds from BEBC
    \cite{Cooper-Sarkar:1981bam,Lyons:1981xs,ParticleDataGroup:2022pth}
    and SBND-PRISM \cite{Alves:2024djc}.
  }
  \label{tab:LFV}
  \begin{center}
    \begin{tabular}{cc}
      \hline Experiment (Uncertainty) & $\mathrm{BR}\left(\pi^{+} \rightarrow \mu^{+} \nu_{\mathrm{e}}\right)$ \\
      \hline\hline BEBC & $8 \times 10^{-3}$ \\
      \hline SBND $(10 \%)$ & $1.5 \times 10^{-3}$ \\
      \hline SBND-PRISM $(10 \%, 5 \%)$ & $1.2 \times 10^{-3}$ \\
      \hline SBND-PRISM $(10 \%, 2 \%)$ & $8.9 \times 10^{-4}$ \\
      \hline nuSTORM$(1 \%)$ & $7.1 \times 10^{-4}$ \\
      \hline Statistics only & $4.7 \times 10^{-5}$ \\
      \hline
    \end{tabular}
  \end{center}
\end{table}    

nuSTORM’s combination of a high‑intensity neutrino beam and a
100‑tonne magnetised liquid‑argon detector provides an optimal
environment for precision studies of rare Standard Model processes.
A paradigmatic example is neutrino trident scattering
\cite{Czyz:1964zz,Lovseth:1971vv,Fujikawa:1971nx,Koike:1971tu}, in
which a (anti)neutrino coherently scatters off a nuclear target
\(\mathcal{H}\), producing a charged lepton pair:
$\overset{(-)}{\nu_\alpha} + \mathcal{H} \;\rightarrow\; \overset{(-)}{\nu_\alpha} + \ell^-_\beta + \ell^+_\gamma + \mathcal{H}\,$,
with \(\{\alpha,\beta,\gamma\}\in\{e,\mu,\tau\}\).
At nuSTORM’s energy regime, this process is dominated by coherent
interactions mediated by both \(W\) and \(Z\) bosons. To date, only
the muon‑pair channel \(\nu_\mu + \mathcal{H} \to \nu_\mu + \mu^+ +
\mu^- + \mathcal{H}\) has been observed, first by CHARM II
\cite{CHARM-II:1990dvf} and subsequently by CCFR \cite{CCFR:1991lpl}
and NuTeV \cite{NuTeV:1998khj}.
While earlier theoretical treatments employed four‑fermion effective
interactions \cite{Czyz:1964zz,Lovseth:1971vv,Fujikawa:1971nx},
Standard Model boson exchange \cite{Brown:1971}, or the equivalent
photon approximation \cite{Altmannshofer:2014,Magill:2017}, we adopt
the full \(2\to4\) cross‑section calculation of
reference~\cite{Ballett:2018uuc}. 
The dominant source of  \(\nu_\mu\) at nuSTORM is pion decay, which
produces a flux approximately two orders of magnitude greater than
that from kaon decay (cf figure~12 of \cite{Adey:2015iha}).
Consequently, event‑rate estimates focus exclusively on the pion
component.
As shown in Table~\ref{Table: Trident Number of events}, nuSTORM’s
projected trident event yield exceeds those of all existing facilities
except the DUNE experiment due to its higher neutrino beam energy.
Potential backgrounds include rare meson decays ($e.g.$
\ \(K^\pm\to\mu^+\mu^-\pi^\pm\)) and other multi‑lepton production
processes.
These backgrounds occur at negligible rates or possess kinematic
signatures distinguishable from the coherent trident topology.
Therefore, the main experimental challenge arises from particle
misidentification, which nuSTORM’s high-resolution detector is
specifically designed to minimise.  
\begin{table}[h]
  \caption{
    Comparison of the total number of events (top row: Coherent,
    bottom row: diffractive) for $\nu$STORM against other experiments
    \cite{Ballett:2018uuc}.
    For DUNE, the numbers in parenthesis signify the antineutrino
    running mode. Calculations are made based on statistics only,
    $i.e.$ no systematic uncertainties.
  }
  \label{Table: Trident Number of events}
  \begin{center}
    \begin{tabular}{cccccc}
      \hline Channel & SBND & $\mu$BooNe & ICARUS & DUNE  & nuSTORM\\
      \hline\hline \multirow[t]{2}{*}{ $\mathrm{e}^{ \pm} \mu^{\mp}$}& 10 & 0.7 & 1 & 2993 (2307)  & 173\\
        & 2 & 0.1 & 0.2 & 692 (530) & 29 \\
      \hline \multirow[t]{2}{*}{ $e^{+} e^{-}$}& 6 & 0.4 & 0.7 & 1007 (800) & 107 \\
        & 0.7 & 0 & 0.1 & 143 (111) &   5\\
      \hline \multirow[t]{2}{*}{ $\mu^{+} \mu^{-}$}& 0.4 & 0 & 0.0 & 286 (210) & 14\\
         
       & 0.4 & 0 & 0.0 & 196 (147) & 9\\
      \hline
    \end{tabular}
  \end{center}
\end{table}

\subsection{Technology test bed}
\label{SubSect:SciCon:TechTstBd}

A Muon Collider has the potential to deliver multi-TeV
lepton-antilepton collisions at a cost and on a timescale advantageous
compared to electron-positron or next-generation hadron
colliders~\cite{accettura2023towards}.
Unlike protons, muons are fundamental particles that can provide
comparable physics outcomes at much lower center-of-mass energies.
The large muon mass leads to reduced synchrotron radiation losses,
allowing for the TeV-scale energies to be reached in a more compact
machine compared to an electron-positron collider.
The International Muon Collider collaboration (IMCC) is developing conceptual designs for facilities capable of operation at centre-of-mass
energies of 3\,TeV and 10\,TeV, which could deliver an integrated
luminosity of $10\,{\rm ab}^{-1}$~\cite{accettura2024interim}. 

One of the key challenges for the Muon Collider development is
delivering a high-brightness muon beam, which is essential for
achieving the target luminosity.
The technique proposed to accomplish this is ionisation cooling,
through which the phase-space volume (emittance) of the muon beam
decreases as the beam traverses an energy-absorbing material.
The principle of ionisation cooling was demonstrated by the Muon
Ionisation Cooling Experiment (MICE)
collaboration~\cite{MICE:Nature}.
MICE determined the reduction in the transverse
emittance of a muon beam as it passed through a single liquid-hydrogen
or lithium-hydride absorber~\cite{mice2024transverse}.
The IMCC has identified a Muon Cooling Demonstrator programme~\cite{accettura2023towards, accettura2024interim} critical
to prove that the technology required to achieve such intense muon
beams using six-dimensional ionisation cooling can be developed,
built, and reliably operated. 

nuSTORM will provide the world's highest power stored muon beam and
will be uniquely well placed to serve as a R\&D platform for muon
accelerator technologies essential for the realisation of a Muon
Collider.
Since nuSTORM and the Muon Collider employ the same pion production
mechanism, a low-energy muon beam can be produced through the capture
of an appropriate pion-beam phase space at the nuSTORM target and
delivered to the Muon Cooling Demonstrator facility.
In addition, the design of the storage ring arcs and return straight
is based on Fixed Field Alternating gradient (FFA) magnets.
Hence, nuSTORM can be used to develop and test the FFA magnet
technology, which has applications for muon fast acceleration and is
an attractive option for the Muon Collider high-energy accelerating
complex.
Finally, nuSTORM will provide a well-characterized muon beam, making
it an ideal testbed for developing and validating beam monitoring
instrumentation required for a Muon Collider.

\graphicspath{ {01-Objectives/Figures} }

\section{Objectives}
\label{Sect:Objctvs}

nuSTORM is based on a low-energy muon decay ring.
Pions, produced in the bombardment of a target, are captured in a
magnetic channel.
The magnetic channel is designed to deliver a pion beam with
momentum $p_\pi$ and momentum spread $\sim \pm 10\%\,p_\pi$ to the 
muon decay ring.
The pion beam is injected into the production straight of the decay
ring.
Depending on energy, between 40\% and 90\% of pions decay 
as the beam passes through the
production straight.
At the end of the straight, the return arc selects a muon beam 
of momentum $p_\mu < p_\pi$ and momentum spread $\sim \pm 16\%\,p_\mu$ 
that then circulates.
Undecayed pions and muons outside the momentum acceptance of the ring
are directed to a beam dump.

A detector placed on the axis of the nuSTORM production straight will
receive a bright flash of muon neutrinos from pion decay followed by a
series of pulses of muon and electron neutrinos from subsequent turns
of the muon beam.
Appropriate instrumentation in the decay ring and production
straight will be capable of determining the integrated neutrino flux
with a precision of $\lsim 1$\%.
The flavour composition of the neutrino beam from muon decay is
known and the neutrino-energy spectrum can be calculated precisely
using the Michel parameters and the optics of the muon decay ring.

\graphicspath{ {01-Methodology/Figures} }

\section{Methodology}
\label{Sect:Mthdgy}

\subsection{Facility overview}
\label{SubSect:Mthdgy:Ovrvw}

\graphicspath { {03-Methodology/03-01-FacilityOverview/Figures/} }

\noindent
The following sections give a brief description of the design
developments of nuSTORM's accelerator complex.
The authors are actively developing the accelerator design, the
detector concept, and the analysis framework.

\subsubsection{Target system and pion transport}

The current design of the pion production and capture system assumes a
CERN site for the facility, with protons extracted from the CERN
Super Proton Synchrotron (SPS).
The 100\,GeV proton beam is focused and impinges on an inconel target
housed inside a focusing horn.
The pions collected by the horn are directed into a short transfer
line with a large momentum acceptance of $\sim\,\pm 10\%$.
The transfer line is composed of dipoles, quadrupoles and collimators,
and transports the pions to the storage ring.  

FLUKA~\cite{Battistoni:2015epi,Ahdida:2022gjl} simulations of the target and horn design were
carried out to characterise the horn pion capture efficiency.
Ongoing studies are aimed at improving the production and capture
yield of low-energy pions ($p_\pi$ $\leq$ 2 GeV/c), and investigate
different horn geometries, including the option of using a two-horn
configuration.
The transfer line design and momentum acceptance have been validated
in tracking studies using the Beam Delivery Simulation software
(BDSIM)~\cite{Nevay:2021wtg,AlvesThesis}. 

\subsubsection{Storage Ring}

\begin{figure}[h]
  \begin{center}
    \begin{overpic}[width=0.8\textwidth]{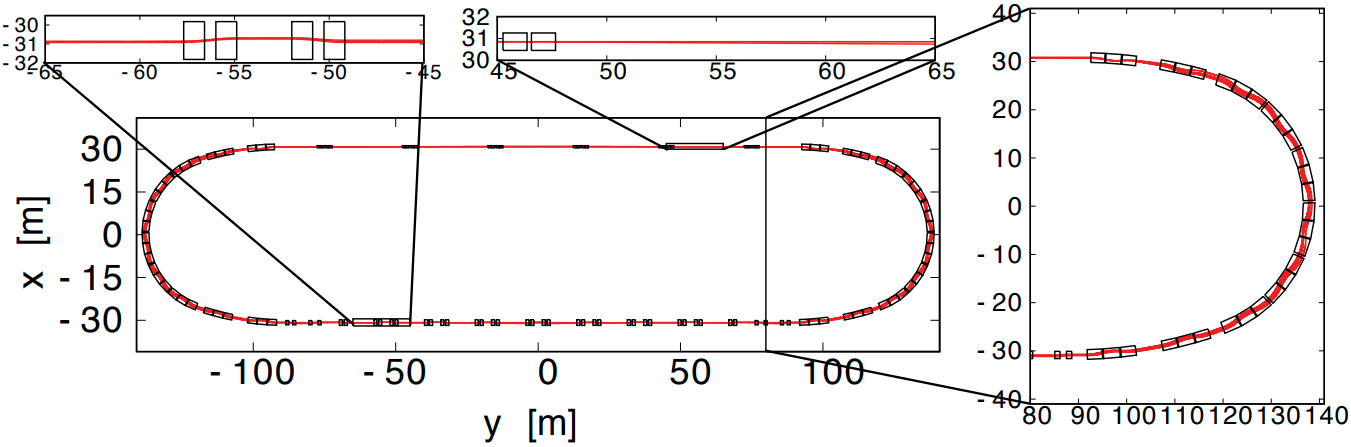}
      \put(29, 19.5){\small Production straight}
      \put(32, 10.5){\small  Return straight}
    \end{overpic}
  \end{center}
  \caption{
    Schematic of the nuSTORM storage ring lattice~\cite{Lagrange:2018rt}.
  }
  \label{fig:ring}
\end{figure}

The nuSTORM decay ring, shown in figure~\ref{fig:ring}, is a compact
racetrack-shaped storage ring $\sim$ 616 m in circumference, composed
of large aperture magnets.
It has been designed to store muon beams with momentum in the 1--6
GeV/c range, with a momentum acceptance of up to $\pm 16 \%$ and
dynamical acceptance of 1 mm rad~\cite{Lagrange:2018rt}.
The arcs and the return straight are based on Fixed Field Alternating
gradient (FFA) magnets to achieve a large dynamic acceptance, while a
conventional focusing-defocusing (FODO) optics is employed in the
production straight to maximise the muon production efficiency.
A model of the production straight has been implemented in BDSIM and
it is used for tracking-based muon production studies.
In addition, active BDSIM development work is aimed at modelling FFA
magnets, to enable tracking-based neutrino-production studies
utilising a model of the entire ring. 

\subsection{Fluxes}
\label{SubSect:Mthdlgy:Flx}

\graphicspath { {03-Methodology/03-02-Fluxes/Figures/} }

\begin{figure}[!h]
  \begin{center}
    \begin{minipage}{0.32\textwidth}
      \begin{center}
        \includegraphics[width=\textwidth]{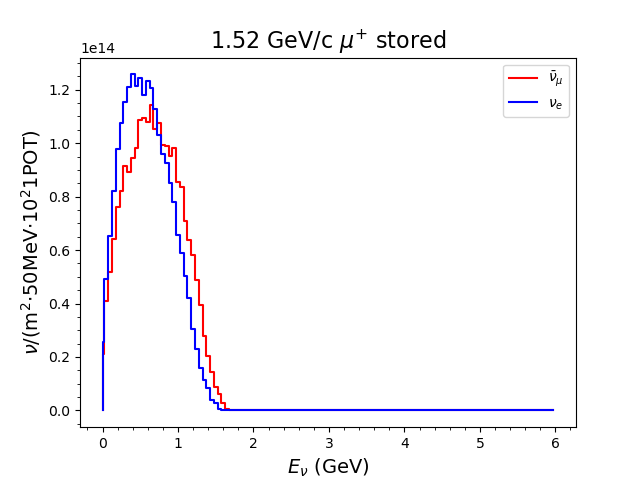}
      \end{center}
    \end{minipage}%
    \hfill
    \begin{minipage}{0.32\textwidth}
      \begin{center}
        \includegraphics[width=\textwidth]{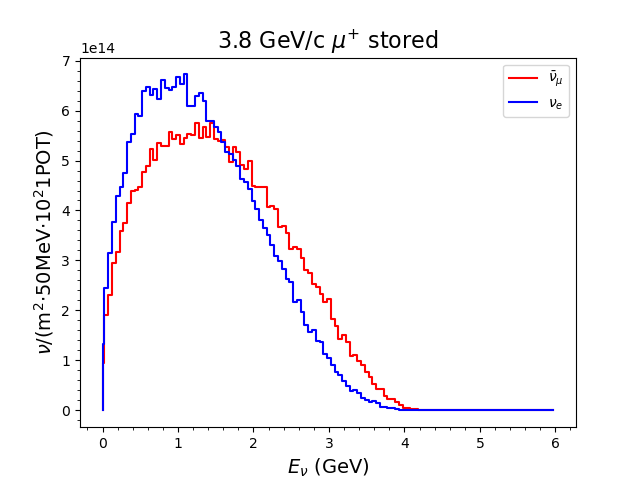}
      \end{center}
    \end{minipage}%
    \hfill
    \begin{minipage}{0.32\textwidth}
      \begin{center}
        \includegraphics[width=\textwidth]{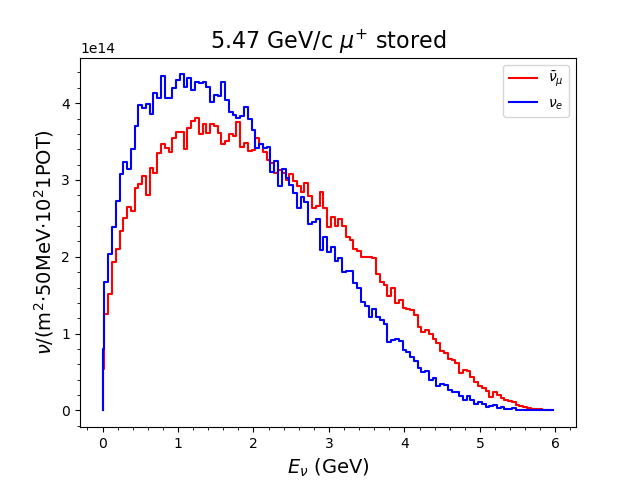}
      \end{center}
    \end{minipage}
  \end{center}
  \caption{Neutrino flux distributions for different stored muon momenta.}
  \label{fig:nuFluxGrid}
\end{figure}

\begin{figure}[h]
  \begin{center}
    \begin{minipage}{0.3\textwidth}
      \begin{center}
        \includegraphics[width=\textwidth]{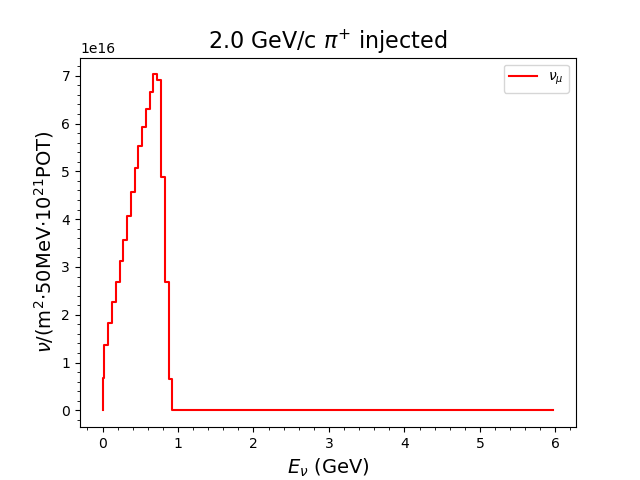}
          \( \nu \) Flux (Pi Flash) for the 1.52 GeV/c muon momentum setting.
      \end{center}
    \end{minipage}%
    \hfill
    \begin{minipage}{0.3\textwidth}
      \centering
      \includegraphics[width=\textwidth]{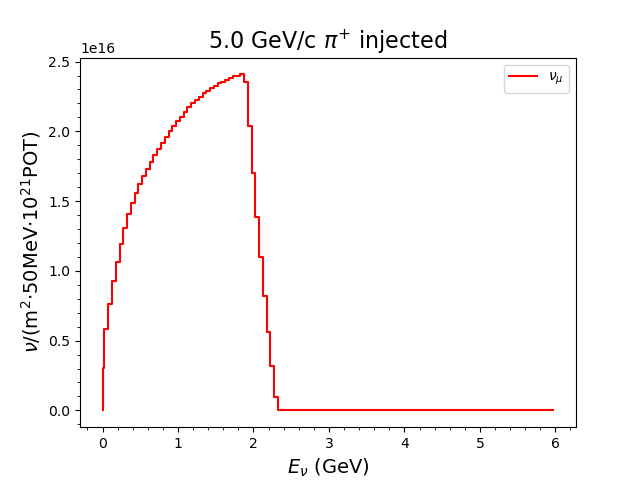}
      \( \nu \) Flux (Pi Flash) for the 3.80 GeV/c muon momentum setting.
    \end{minipage}%
    \hfill
    \begin{minipage}{0.3\textwidth}
      \centering
      \includegraphics[width=\textwidth]{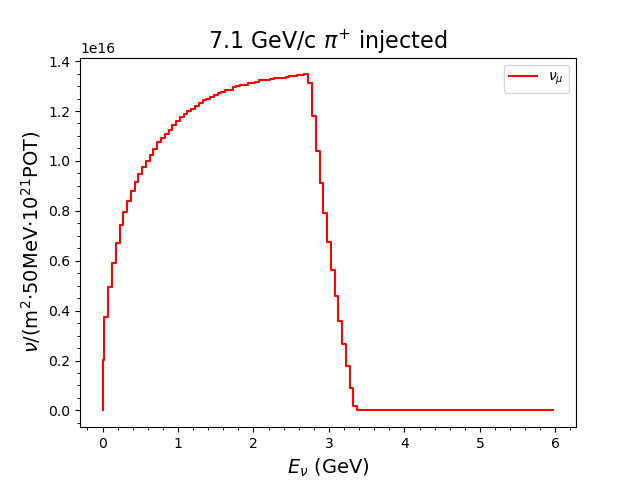}
      \( \nu \) Flux (Pi Flash) for the 5.47 GeV/c muon momentum setting.
    \end{minipage}
  \end{center}
  \caption{
    Neutrino flux distributions for different pion energies
    (Pi Flash). The muons stored from the decay of these pions generate the corresponding fluxes shown in Fig \ref{fig:nuFluxGrid}.
  } 
  \label{fig:nuFluxPiFlashGrid}
\end{figure}

The neutrino spectrum has been generated using NuSIM \cite{nuSTORMsnow}, and the fluxes
normalised to protons on target (POT) are shown in
figure~\ref{fig:nuFluxGrid} for 3 different muon momentum
settings.
The corresponding fluxes from the pion flashes are also shown in
figure~\ref{fig:nuFluxPiFlashGrid}.

\subsection{Detector considerations}
\label{SubSect:Mthdlgy:Dtctr}

The detector suite will provide a comprehensive neutrino-interaction
physics program with exceptional precision and an expansive ability to
search for new phenomena. 
To realise the full potential of the facility requires:
\begin{enumerate}   
    \item {Identification of neutrino flavours through charge detection in a magnet, supplemented by muon and electron differentiation.}
    \item {Ultra-low tracking threshold in $4\pi$ acceptance.}
    \item {Charged hadron identification.}
    \item {Ability to detect neutral particles using calorimetry and time-of-flight measurements.}
    \item {Fast detector capable of self-triggering and handling high event rates.}
    \item A variety of nuclear targets to measure cross-sections as a
    function of the nuclear target mass number (A).
\end{enumerate}

Concepts developed for the near detectors of longbaseline neutrino
oscillation experiments are suitable options for detectors at
nuSTORM.
These include SuperFGD, a highly-segmented tracking scintillator
incorporated in the upgraded T2K near
detector~\cite{Blanchet:2021pqb}, or detectors being developed for the
DUNE near detector complex~\cite{DUNE:2021tad}, including:
\begin{enumerate}[noitemsep,topsep=0pt]
  \item A pixelated LAr detector;
  \item A magnetised high-pressure gaseous TPC (HPgTPC); and
  \item A straw-Tube trackers (STT) with thin targets.
\end{enumerate}
The SuperFGD, HPgTPC, and STT can be magnetised, while
magnetisation concepts for a pixelated LAr detector have been
developed, but are expensive.
However, current R\&D on high-temperature superconductor magnets may
make the magnet system of a LAr detector affordable. 

DUNE's HPgTPC~\cite{DUNE:2021tad} is surrounded by an electromagnetic calorimeter and a
superconducting solenoid.
The 0.5~T magnetic volume is large, approximately 7\,m in diameter and
7.5\,m long (both the HPgTPC and the calorimeter are in the magnetic
volume).
A muon system is located outside the solenoid, in the return iron.
This detector offers many advantages including: capability to vary the
target nucleus (main gas component) from He to Xe, operation at
pressures from 1 bar to 10 bar, 4$\pi$ tracking with track thresholds
down to 5\,MeV, excellent particle ID which allows for very precise
determination of exclusive final states and the addition of a magnetic
field allows for energy measurement via spectrometry as well as
calorimetry (from the ECAL). The drawback is the relatively low mass
(about 1\,tonne fiducial volume for Argon at 10\,bar).
In DUNE, HPgTPC functions as muon catcher for the pixelated LAr
detector which is just upstream.
The return iron has a window which allows muons that exit the LAr to
be accurately momentum analysed in the HPgTPC.  

In addition to these concepts, a recently demonstrated
technique~\cite{LiquidO:2019mxd} using an opaque scintillator and a
dense array of optical fibres, could be considered as it provides
high-resolution imaging and efficient identification of individual
particles.

\graphicspath{ {01-Objectives/Figures} }

\section{Readiness, challenges, timeline and costs}
\label{Sect:RdnssChllngsTmlnCsts}

The implementation of nuSTORM at CERN was considered in the light of
present commitments and likely future developments~\cite{Ahdida:2020whw}.
These considerations indicated that implementation of nuSTORM might be
possible from around 2030.
Such an ambition is consistent with the following considerations:
\begin{itemize}[noitemsep,topsep=0pt]
  \item Upgrades to the injectors and LHC follow the anticipated
    Long Shutdown (LS)/Run timetable;
  \item If approval for the development of a CDR were to be granted,
    and resources assigned, a 3-year comprehensive Design Study would
    need to be carried out.
    An R\&D period of around 3 years is likely to be required once the
    design-study phase is complete.
    The R\&D phase would culminate in the publication of a TDR.
    Should the project then be approved, a period of component
    production and preparation for project execution would be
    required.
\end{itemize}
In consequence, a 10-year programme for the implementation of nuSTORM
is a reasonable aspiration.
The cost was estimated in~\cite{Ahdida:2020whw} to be the region of
156\,MCHF.

\newpage
\bibliographystyle{99-Styles/utphys}
\bibliography{Concatenated-bibliography}

\providecommand{\href}[2]{#2}\begingroup\raggedright\begin{thebibliography}{10}

\bibitem{nuSTORMsnow}
L.~A. Ruso, T.~Alves, S.~Boyd, A.~Bross, P.~R. Hobson, P.~Kyberd, J.~B. Lagrange, K.~Long, X.-G. Lu, J.~Pasternak, {\em et al.}, ``Neutrinos from stored muons (nustorm),'' {\em arXiv preprint arXiv:2203.07545} (2022) , \href{http://arxiv.org/abs/2203.07545}{{\ttfamily arXiv:2203.07545 [hep-ex]}}.

\bibitem{Ahdida:2020whw}
C.~C. Ahdida {\em et al.}, ``{nuSTORM at CERN: Feasibility Study},'' \href{http://dx.doi.org/10.17181/CERN.FQTB.O8QN}{{\em CERN-PBC-REPORT-2019-003} (10, 2020) }.

\bibitem{Hyper-KamiokandeProto-:2015xww}
{\bfseries Hyper-Kamiokande Proto-} Collaboration, K.~Abe {\em et al.}, ``{Physics potential of a long-baseline neutrino oscillation experiment using a J-PARC neutrino beam and Hyper-Kamiokande},'' \href{http://dx.doi.org/10.1093/ptep/ptv061}{{\em PTEP} {\bfseries 2015} (2015) 053C02}, \href{http://arxiv.org/abs/1502.05199}{{\ttfamily arXiv:1502.05199 [hep-ex]}}.

\bibitem{Lu:2015tcr}
X.~G. Lu, L.~Pickering, S.~Dolan, G.~Barr, D.~Coplowe, Y.~Uchida, D.~Wark, M.~O. Wascko, A.~Weber, and T.~Yuan, ``{Measurement of nuclear effects in neutrino interactions with minimal dependence on neutrino energy},'' \href{http://dx.doi.org/10.1103/PhysRevC.94.015503}{{\em Phys. Rev. C} {\bfseries 94} no.~1, (2016) 015503}, \href{http://arxiv.org/abs/1512.05748}{{\ttfamily arXiv:1512.05748 [nucl-th]}}.

\bibitem{Buss:2011mx}
O.~Buss, T.~Gaitanos, K.~Gallmeister, H.~van Hees, M.~Kaskulov, O.~Lalakulich, A.~B. Larionov, T.~Leitner, J.~Weil, and U.~Mosel, ``{Transport-theoretical Description of Nuclear Reactions},'' \href{http://dx.doi.org/10.1016/j.physrep.2011.12.001}{{\em Phys. Rept.} {\bfseries 512} (2012) 1--124}, \href{http://arxiv.org/abs/1106.1344}{{\ttfamily arXiv:1106.1344 [hep-ph]}}.

\bibitem{Mosel:2023zek}
U.~Mosel and K.~Gallmeister, ``{Lepton-induced reactions on nuclei in a wide kinematical regime},'' \href{http://dx.doi.org/10.1103/PhysRevD.109.033008}{{\em Phys. Rev. D} {\bfseries 109} no.~3, (2024) 033008}, \href{http://arxiv.org/abs/2308.16161}{{\ttfamily arXiv:2308.16161 [nucl-th]}}.

\bibitem{DUNE:2021tad}
{\bfseries DUNE} Collaboration, V.~Hewes {\em et al.}, ``{Deep Underground Neutrino Experiment (DUNE) Near Detector Conceptual Design Report},'' \href{http://dx.doi.org/10.3390/instruments5040031}{{\em Instruments} {\bfseries 5} no.~4, (2021) 31}, \href{http://arxiv.org/abs/2103.13910}{{\ttfamily arXiv:2103.13910 [physics.ins-det]}}.

\bibitem{Bhadra2014LetterBeamlineb}
S.~Bhadra, A.~Blondel, S.~Bordoni, A.~Bravar, C.~Bronner, J.~Caravaca-Rodriguez, M.~Dziewiecki, T.~Feusels, G.~A. Fiorentini-Aguirre, M.~Friend, L.~Haegel, M.~Hartz, R.~Henderson, T.~Ishida, M.~Ishitsuka, C.~K. Jung, A.~C. Kaboth, H.~Kakuno, H.~Kamano, A.~Konaka, Y.~Kudenko, M.~Kuze, T.~Lindner, K.~Mahn, J.~F. Martin, J.~Marzec, K.~S. McFarland, S.~Nakayama, T.~Nakaya, S.~Nakamura, Y.~Nishimura, A.~Rychter, F.~Sanchez, T.~Sato, M.~Scott, T.~Sekiguchi, M.~Shiozawa, T.~Sumiyoshi, R.~Tacik, H.~K. Tanaka, H.~A. Tanaka, S.~Tobayama, M.~Vagins, J.~Vo, D.~Wark, M.~O. Wascko, M.~J. Wilking, S.~Yen, M.~Yokoyama, and M.~Ziembicki, ``{Letter of Intent to Construct a nuPRISM Detector in the J-PARC Neutrino Beamline},''. \url{https://arxiv.org/abs/1412.3086v2}.

\bibitem{Kamath2024SimulationExperiment}
R.~Kamath, ``{Simulation and study of the nuSTORM (neutrinos from Stored Muons) experiment},'' {\em JACoW} {\bfseries IPAC2024} (2024) TUAD3.

\bibitem{Shanahan2021PhysicsReview}
P.~Shanahan and P.~Vahle, ``{Physics with NOvA: a half-time review},'' \href{http://dx.doi.org/10.1140/epjs/s11734-021-00285-9}{{\em European Physical Journal: Special Topics} {\bfseries 230} no.~24, (2021) }.

\bibitem{Chakraborty_2021}
K.~Chakraborty, S.~Goswami, and K.~Long, ``New physics at nustorm,'' \href{http://dx.doi.org/10.1103/physrevd.103.075009}{{\em Physical Review D} {\bfseries 103} no.~7, (Apr., 2021) }. \url{http://dx.doi.org/10.1103/PhysRevD.103.075009}.

\bibitem{Arkani-Hamed:1998jmv}
N.~Arkani-Hamed, S.~Dimopoulos, and G.~R. Dvali, ``{The Hierarchy problem and new dimensions at a millimeter},'' \href{http://dx.doi.org/10.1016/S0370-2693(98)00466-3}{{\em Phys. Lett. B} {\bfseries 429} (1998) 263--272}, \href{http://arxiv.org/abs/hep-ph/9803315}{{\ttfamily arXiv:hep-ph/9803315}}.

\bibitem{Arkani-Hamed:1998sfv}
N.~Arkani-Hamed, S.~Dimopoulos, and G.~R. Dvali, ``{Phenomenology, astrophysics and cosmology of theories with submillimeter dimensions and TeV scale quantum gravity},'' \href{http://dx.doi.org/10.1103/PhysRevD.59.086004}{{\em Phys. Rev. D} {\bfseries 59} (1999) 086004}, \href{http://arxiv.org/abs/hep-ph/9807344}{{\ttfamily arXiv:hep-ph/9807344}}.

\bibitem{Penedo:2022etl}
J.~T. Penedo and J.~a. Pulido, ``{Baseline and other effects for a sterile neutrino at DUNE},'' \href{http://dx.doi.org/10.1103/PhysRevD.107.075026}{{\em Phys. Rev. D} {\bfseries 107} no.~7, (2023) 075026}, \href{http://arxiv.org/abs/2207.02331}{{\ttfamily arXiv:2207.02331 [hep-ph]}}.

\bibitem{Kopp:2013vaa}
J.~Kopp, P.~A.~N. Machado, M.~Maltoni, and T.~Schwetz, ``{Sterile Neutrino Oscillations: The Global Picture},'' \href{http://dx.doi.org/10.1007/JHEP05(2013)050}{{\em JHEP} {\bfseries 05} (2013) 050}, \href{http://arxiv.org/abs/1303.3011}{{\ttfamily arXiv:1303.3011 [hep-ph]}}.

\bibitem{Declais:1994su}
Y.~Declais {\em et al.}, ``Search for neutrino oscillations at 15-meters, 40-meters, and 95-meters from a nuclear power reactor at bugey,''
\href{http://dx.doi.org/10.1016/0550-3213(94)00513-E}{{\em Nucl. Phys. B} {\bfseries 434} (1995) 503--534}.

\bibitem{Declais:1994ma}
Y.~Declais, H.~de~Kerret, B.~Lefievre, M.~Obolensky, A.~Etenko, {\em et al.}, ``{Study of reactor anti-neutrino interaction with proton at Bugey nuclear power plant},'' \href{http://dx.doi.org/10.1016/0370-2693(94)91394-3}{{\em Phys.Lett.} {\bfseries B338} (1994) 383--389}.

\bibitem{CALTECH-SIN-TUM:1986xvg}
{\bfseries CALTECH-SIN-TUM} Collaboration, G.~Zacek {\em et al.}, ``{Neutrino Oscillation Experiments at the Gosgen Nuclear Power Reactor},'' \href{http://dx.doi.org/10.1103/PhysRevD.34.2621}{{\em Phys. Rev. D} {\bfseries 34} (1986) 2621--2636}.

\bibitem{Kwon:1981ua}
H.~Kwon, F.~Boehm, A.~Hahn, H.~Henrikson, J.~Vuilleumier, {\em et al.}, ``{Search for neutrino oscillations at a fission reactor},'' \href{http://dx.doi.org/10.1103/PhysRevD.24.1097}{{\em Phys.Rev.} {\bfseries D24} (1981) 1097--1111}.

\bibitem{Vidyakin:1987ue}
G.~Vidyakin, V.~Vyrodov, I.~Gurevich, Y.~Kozlov, V.~Martemyanov, {\em et al.}, ``{DETECTION OF ANTI-NEUTRINOS IN THE FLUX FROM TWO REACTORS},'' {\em Sov.Phys.JETP} {\bfseries 66} (1987) 243--247.

\bibitem{Greenwood:1996pb}
Z.~D. Greenwood {\em et al.}, ``{Results of a two position reactor neutrino oscillation experiment},'' \href{http://dx.doi.org/10.1103/PhysRevD.53.6054}{{\em Phys. Rev. D} {\bfseries 53} (1996) 6054--6064}.

\bibitem{Kuvshinnikov:1990ry}
A.~Kuvshinnikov, L.~Mikaelyan, S.~Nikolaev, M.~Skorokhvatov, and A.~Etenko, ``{Measuring the anti-electron-neutrino + p -> n + e+ cross-section and beta decay axial constant in a new experiment at Rovno NPP reactor. (In Russian)},'' {\em JETP Lett.} {\bfseries 54} (1991) 253--257.

\bibitem{Afonin:1988gx}
A.~I. Afonin, S.~N. Ketov, V.~I. Kopeikin, L.~A. Mikaelyan, M.~D. Skorokhvatov, and S.~V. Tolokonnikov, ``{A Study of the Reaction $\bar{\nu}_e + P \to e^+ + N$ on a Nuclear Reactor},'' {\em Sov. Phys. JETP} {\bfseries 67} (1988) 213--221.

\bibitem{MINOS:2011ysd}
{\bfseries MINOS} Collaboration, P.~Adamson {\em et al.}, ``{Active to sterile neutrino mixing limits from neutral-current interactions in MINOS},'' \href{http://dx.doi.org/10.1103/PhysRevLett.107.011802}{{\em Phys. Rev. Lett.} {\bfseries 107} (2011) 011802}, \href{http://arxiv.org/abs/1104.3922}{{\ttfamily arXiv:1104.3922 [hep-ex]}}.

\bibitem{Dydak:1983zq}
F.~Dydak {\em et al.}, ``{A Search for Muon-neutrino Oscillations in the $\Delta m^2$ Range 0.3-eV$^2$ to 90-eV$^2$},''
\href{http://dx.doi.org/10.1016/0370-2693(84)90688-9}{{\em Phys. Lett. B} {\bfseries 134} (1984) 281}.

\bibitem{Bilenky:1999ny}
S.~M. Bilenky, C.~Giunti, W.~Grimus, and T.~Schwetz, ``{Four-neutrino mass spectra and the Super-Kamiokande atmospheric up-down asymmetry},'' \href{http://dx.doi.org/10.1103/PhysRevD.60.073007}{{\em Phys. Rev. D} {\bfseries 60} (1999) 073007},
\href{http://arxiv.org/abs/9903454}{{\ttfamily arXiv:9903454 [hep-ph]}}.

\bibitem{MiniBooNE:2012meu}
{\bfseries MiniBooNE, SciBooNE} Collaboration, G.~Cheng {\em et al.}, ``{Dual baseline search for muon antineutrino disappearance at $0.1 {\rm eV}^2 < {\Delta}m^2 < 100 {\rm eV}^2$},'' \href{http://dx.doi.org/10.1103/PhysRevD.86.052009}{{\em Phys. Rev. D} {\bfseries 86} (2012) 052009}, \href{http://arxiv.org/abs/1208.0322}{{\ttfamily arXiv:1208.0322 [hep-ex]}}.

\bibitem{Kuno:1999jp}
Y.~Kuno and Y.~Okada, ``Muon decay and physics beyond the standard model,'' {\em Rev. Mod. Phys.} {\bfseries 73} (2001) 151--202,
\href{http://arxiv.org/abs/hep-ph/9909265}{{\ttfamily hep-ph/9909265}}.

\bibitem{Bernstein:2013hba}
R.~H. Bernstein and P.~S. Cooper, ``{Charged Lepton Flavor Violation: An Experimenter's Guide},'' \href{http://dx.doi.org/10.1016/j.physrep.2013.07.002}{{\em Phys. Rept.} {\bfseries 532} (2013) 27--64}, \href{http://arxiv.org/abs/1307.5787}{{\ttfamily arXiv:1307.5787 [hep-ex]}}.

\bibitem{Lindner:2016bgg}
M.~Lindner, M.~Platscher, and F.~S. Queiroz, ``A call for new physics: The muon anomalous magnetic moment and lepton flavor violation,'' {\em Phys. Rept.} {\bfseries 731} (2018) 1--82, \href{http://arxiv.org/abs/1610.06587}{{\ttfamily arXiv:1610.06587 [hep-ph]}}.

\bibitem{Calibbi:2017uvl}
L.~Calibbi and G.~Signorelli, ``Charged lepton flavour violation: An experimental and theoretical introduction,'' {\em Riv. Nuovo Cim.} {\bfseries 41} (2018) 71--174, \href{http://arxiv.org/abs/1709.00294}{{\ttfamily arXiv:1709.00294 [hep-ph]}}.

\bibitem{MEG:2016leq}
{\bfseries MEG} Collaboration, A.~M. Baldini {\em et al.}, ``{Search for the lepton flavour violating decay $\mu ^+ \rightarrow \mathrm {e}^+ \gamma $ with the full dataset of the MEG experiment},'' \href{http://dx.doi.org/10.1140/epjc/s10052-016-4271-x}{{\em Eur. Phys. J. C} {\bfseries 76} no.~8, (2016) 434}, \href{http://arxiv.org/abs/1605.05081}{{\ttfamily arXiv:1605.05081 [hep-ex]}}.

\bibitem{Bernstein:2019fyh}
{\bfseries Mu2e} Collaboration, R.~H. Bernstein, ``{The Mu2e Experiment},'' \href{http://dx.doi.org/10.3389/fphy.2019.00001}{{\em Front. in Phys.} {\bfseries 7} (2019) 1}, \href{http://arxiv.org/abs/1901.11099}{{\ttfamily arXiv:1901.11099 [physics.ins-det]}}.

\bibitem{Cooper-Sarkar:1981bam}
A.~M. Cooper-Sarkar, J.~G. Guy, A.~G. Michette, M.~Tyndel, and W.~Venus, ``{Limits on Neutrino - Anti-neutrinos Transitions From a Study of High-energy Neutrino Interactions},'' \href{http://dx.doi.org/10.1016/0370-2693(82)90914-5}{{\em Phys. Lett. B} {\bfseries 112} (1982) 97--99}.

\bibitem{Lyons:1981xs}
L.~Lyons, C.~Albajar, and G.~Myatt, ``{An Experimental Limit on the Decay $K^+ \to \mu^+ \nu_e$},'' \href{http://dx.doi.org/10.1007/BF01549729}{{\em Z. Phys. C} {\bfseries 10} (1981) 215}.

\bibitem{ParticleDataGroup:2022pth}
{\bfseries Particle Data Group} Collaboration, R.~L. Workman {\em et al.}, ``{Review of Particle Physics},'' \href{http://dx.doi.org/10.1093/ptep/ptac097}{{\em PTEP} {\bfseries 2022} (2022) 083C01}.

\bibitem{Alves:2024djc}
G.~F.~S. Alves, R.~Z. Funchal, and P.~A.~N. Machado, ``{Could SBND-PRISM probe lepton flavor violation?},'' \href{http://dx.doi.org/10.1103/PhysRevD.110.035031}{{\em Phys. Rev. D} {\bfseries 110} no.~3, (2024) 035031}, \href{http://arxiv.org/abs/2405.00777}{{\ttfamily arXiv:2405.00777 [hep-ph]}}.

\bibitem{Czyz:1964zz}
W.~Czyz, G.~C. Sheppey, and J.~D. Walecka, ``{Neutrino production of lepton pairs through the point four-fermion interaction},'' \href{http://dx.doi.org/10.1007/BF02734586}{{\em Nuovo Cim.} {\bfseries 34} (1964) 404--435}.

\bibitem{Lovseth:1971vv}
J.~Lovseth and M.~Radomiski, ``{Kinematical distributions of neutrino-produced lepton triplets},'' \href{http://dx.doi.org/10.1103/PhysRevD.3.2686}{{\em Phys. Rev. D} {\bfseries 3} (1971) 2686--2706}.

\bibitem{Fujikawa:1971nx}
K.~Fujikawa, ``{The self-coupling of weak lepton currents in high-energy neutrino and muon reactions},'' \href{http://dx.doi.org/10.1016/0003-4916(71)90244-2}{{\em Annals Phys.} {\bfseries 68} (1971) 102--162}.

\bibitem{Koike:1971tu}
K.~Koike, M.~Konuma, K.~Kurata, and K.~Sugano, ``{Neutrino production of lepton pairs. 1. -},'' \href{http://dx.doi.org/10.1143/PTP.46.1150}{{\em Prog. Theor. Phys.} {\bfseries 46} (1971) 1150--1169}.

\bibitem{CHARM-II:1990dvf}
{\bfseries CHARM-II} Collaboration, D.~Geiregat {\em et al.}, ``{First observation of neutrino trident production},'' \href{http://dx.doi.org/10.1016/0370-2693(90)90146-W}{{\em Phys. Lett. B} {\bfseries 245} (1990) 271--275}.

\bibitem{CCFR:1991lpl}
{\bfseries CCFR} Collaboration, S.~R. Mishra {\em et al.}, ``{Neutrino Tridents and W Z Interference},'' \href{http://dx.doi.org/10.1103/PhysRevLett.66.3117}{{\em Phys. Rev. Lett.} {\bfseries 66} (1991) 3117--3120}.

\bibitem{NuTeV:1998khj}
{\bfseries NuTeV} Collaboration, T.~Adams {\em et al.}, ``{Neutrino trident production from NuTeV},'' in {\em {29th International Conference on High-Energy Physics}}, pp.~631--634.
\newblock 7, 1998.
\newblock \href{http://arxiv.org/abs/hep-ex/9811012}{{\ttfamily arXiv:hep-ex/9811012}}.

\bibitem{Brown:1971}
R.~W. Brown, ``Intermediate boson. i. theoretical production cross-sections in high-energy neutrino and muon experiments,'' {\em Phys. Rev. D} {\bfseries 3} (1971) 207.

\bibitem{Altmannshofer:2014}
M.~P. W.~Altmannshofer, S.~Gori and I.~Yavin, ``Neutrino trident production: A powerful probe of new physics with neutrino beams,'' {\em Phys. Rev. Lett.} {\bfseries 113} (2014) 091801, \href{http://arxiv.org/abs/1406.2332}{{\ttfamily arXiv:1406.2332}}.

\bibitem{Magill:2017}
G.~Magill and R.~Plestid, ``Neutrino trident production at the intensity frontier,'' {\em Phys. Rev. D} {\bfseries 95} (2017) 073004, \href{http://arxiv.org/abs/1612.05642}{{\ttfamily arXiv:1612.05642}}.

\bibitem{Ballett:2018uuc}
P.~Ballett, M.~Hostert, S.~Pascoli, Y.~F. Perez-Gonzalez, Z.~Tabrizi, and R.~Zukanovich~Funchal, ``{Neutrino Trident Scattering at Near Detectors},'' \href{http://dx.doi.org/10.1007/JHEP01(2019)119}{{\em JHEP} {\bfseries 01} (2019) 119}, \href{http://arxiv.org/abs/1807.10973}{{\ttfamily arXiv:1807.10973 [hep-ph]}}.

\bibitem{Adey:2015iha}
D.~Adey, R.~Bayes, A.~Bross, and P.~Snopok, ``{nuSTORM and A Path to a Muon Collider},'' \href{http://dx.doi.org/10.1146/annurev-nucl-102014-021930}{{\em Ann. Rev. Nucl. Part. Sci.} {\bfseries 65} (2015) 145--175}.

\bibitem{accettura2023towards}
C.~Accettura, D.~Adams, R.~Agarwal, C.~Ahdida, C.~Aim{\`e}, N.~Amapane, D.~Amorim, P.~Andreetto, F.~Anulli, R.~Appleby, {\em et al.}, ``Towards a muon collider,'' \href{http://dx.doi.org/https://doi.org/10.1140/epjc/s10052-023-11889-x}{{\em Eur. Phys. J. C} {\bfseries 83} no.~9, (2023) 864}.

\bibitem{accettura2024interim}
C.~Accettura, S.~Adrian, R.~Agarwal, C.~Ahdida, C.~Aim{\'e}, A.~Aksoy, G.~Alberghi, S.~Alden, N.~Amapane, D.~Amorim, {\em et al.}, ``Interim report for the international muon collider collaboration (imcc),'' {\em arXiv preprint arXiv:2407.12450} (2024) , \href{http://arxiv.org/abs/2407.12450}{{\ttfamily arXiv:2407.12450 [physics.acc-ph]}}.

\bibitem{MICE:Nature}
M.~Bogomilov, R.~Tsenov, G.~Vankova-Kirilova, Y.~P. Song, J.~Y. Tang, Z.~H. Li, R.~Bertoni, M.~Bonesini, F.~Chignoli, R.~Mazza, V.~Palladino, A.~de~Bari, D.~Orestano, L.~Tortora, Y.~Kuno, H.~Sakamoto, A.~Sato, S.~Ishimoto, M.~Chung, C.~K. Sung, F.~Filthaut, D.~Jokovic, D.~Maletic, M.~Savic, N.~Jovancevic, J.~Nikolov, M.~Vretenar, S.~Ramberger, R.~Asfandiyarov, A.~Blondel, F.~Drielsma, Y.~Karadzhov, S.~Boyd, J.~R. Greis, T.~Lord, C.~Pidcott, I.~Taylor, G.~Charnley, N.~Collomb, K.~Dumbell, A.~Gallagher, A.~Grant, S.~Griffiths, T.~Hartnett, B.~Martlew, A.~Moss, A.~Muir, I.~Mullacrane, A.~Oates, P.~Owens, G.~Stokes, P.~Warburton, C.~White, D.~Adams, V.~Bayliss, J.~Boehm, T.~W. Bradshaw, C.~Brown, M.~Courthold, J.~Govans, M.~Hills, J.~B. Lagrange, C.~Macwaters, A.~Nichols, R.~Preece, S.~Ricciardi, C.~Rogers, T.~Stanley, J.~Tarrant, M.~Tucker, S.~Watson, A.~Wilson, R.~Bayes, J.~C. Nugent, F.~J.~P. Soler, G.~T. Chatzitheodoridis, A.~J. Dick, K.~Ronald, C.~G. Whyte, A.~R. Young, R.~Gamet, P.~Cooke, V.~J. Blackmore,
  D.~Colling, A.~Dobbs, P.~Dornan, P.~Franchini, C.~Hunt, P.~B. Jurj, A.~Kurup, K.~Long, J.~Martyniak, S.~Middleton, J.~Pasternak, M.~A. Uchida, J.~H. Cobb, C.~N. Booth, P.~Hodgson, J.~Langlands, E.~Overton, V.~Pec, P.~J. Smith, S.~Wilbur, M.~Ellis, R.~B.~S. Gardener, P.~Kyberd, J.~J. Nebrensky, A.~DeMello, S.~Gourlay, A.~Lambert, D.~Li, T.~Luo, S.~Prestemon, S.~Virostek, M.~Palmer, H.~Witte, D.~Adey, A.~D. Bross, D.~Bowring, A.~Liu, D.~Neuffer, M.~Popovic, P.~Rubinov, B.~Freemire, P.~Hanlet, D.~M. Kaplan, T.~A. Mohayai, D.~Rajaram, P.~Snopok, Y.~Torun, L.~M. Cremaldi, D.~A. Sanders, D.~J. Summers, L.~R. Coney, G.~G. Hanson, C.~Heidt, and M.~collaboration, ``Demonstration of cooling by the muon ionization cooling experiment,'' \href{http://dx.doi.org/10.1038/s41586-020-1958-9}{{\em Nature} {\bfseries 578} no.~7793, (2020) 53--59}. \url{https://doi.org/10.1038/s41586-020-1958-9}.

\bibitem{mice2024transverse}
``Transverse emittance reduction in muon beams by ionization cooling,'' \href{http://dx.doi.org/https://doi.org/10.1038/s41567-024-02547-4}{{\em Nat. Phys.} {\bfseries 20} (2024) 1558--1563}.

\bibitem{Battistoni:2015epi}
G.~Battistoni {\em et al.}, ``{Overview of the FLUKA code},'' \href{http://dx.doi.org/10.1016/j.anucene.2014.11.007}{{\em Annals Nucl. Energy} {\bfseries 82} (2015) 10--18}.

\bibitem{Ahdida:2022gjl}
C.~Ahdida {\em et al.}, ``{New Capabilities of the FLUKA Multi-Purpose Code},'' \href{http://dx.doi.org/10.3389/fphy.2021.788253}{{\em Front. in Phys.} {\bfseries 9} (2022) 788253}.

\bibitem{Nevay:2021wtg}
L.~Nevay {\em et al.}, ``{Recent BDSIM Related Developments and Modeling of Accelerators},'' \href{http://dx.doi.org/10.18429/JACoW-IPAC2021-THPAB214}{{\em JACoW} {\bfseries IPAC2021} (2021) THPAB214}.

\bibitem{AlvesThesis}
T.~Alves, ``Simulating {B}eamline {O}ptics for nu{STORM},'' {\em MSc thesis, Imperial College London} (2022) . \url{https://www.nustorm.org/trac/raw-attachment/wiki/Communication/TechnicalNotes/2022/nuSTORM-TN-01.pdf}.

\bibitem{Lagrange:2018rt}
J.-B. Lagrange, R.~Appleby, J.~Garland, J.~Pasternak, and S.~Tygier, ``{Racetrack {FFAG} muon decay ring for {nuSTORM} with triplet focusing},'' \href{http://dx.doi.org/10.1088/1748-0221/13/09/p09013}{{\em JINST} {\bfseries 13} no.~9, (2018) P09013}, \href{http://arxiv.org/abs/1806.02172}{{\ttfamily arXiv:1806.02172 [physics.acc-ph]}}.

\bibitem{Blanchet:2021pqb}
A.~Blanchet, ``{Physics and Performance of the Upgraded T2K\textquoteright{}s Near Detector},'' \href{http://dx.doi.org/10.1134/S1063778821040086}{{\em Phys. At. Nucl.} {\bfseries 84} no.~4, (2021) 519--523}.

\bibitem{LiquidO:2019mxd}
{\bfseries LiquidO} Collaboration, A.~Cabrera {\em et al.}, ``{Neutrino Physics with an Opaque Detector},'' \href{http://dx.doi.org/10.1038/s42005-021-00763-5}{{\em Commun. Phys.} {\bfseries 4} (2021) 273}, \href{http://arxiv.org/abs/1908.02859}{{\ttfamily arXiv:1908.02859 [physics.ins-det]}}.

\end{thebibliography}\endgroup

\end{document}